\documentclass[preprint,aps,floatfix,superscriptaddress,pre,showpacs]{revtex4}
\usepackage{amsmath, amsthm, amssymb, graphicx}
\usepackage{color}

\input epsf.sty
\begin{document}

\title{Non-equilibrium statistical mechanics of a two-temperature Ising ring
with conserved dynamics}
\author{Nicholas Borchers}
\affiliation{Department of Physics, Virginia Tech, Blacksburg, VA 24061-0435, USA}
\author{Michel Pleimling}
\affiliation{Department of Physics, Virginia Tech, Blacksburg, VA 24061-0435, USA}
\author{R. K. P. Zia}
\affiliation{Department of Physics, Virginia Tech, Blacksburg, VA 24061-0435, USA}
\affiliation{Department of Physics and Astronomy, Iowa State University, Ames, IA 50011,
USA}
\date{\today}

\begin{abstract}
The statistical mechanics of a one-dimensional Ising model in thermal
equilibrium is well-established, textbook material. Yet, when driven far from
equilibrium by coupling two sectors to two baths at different temperatures,
it exhibits remarkable phenomena,
including an unexpected 'freezing by heating.' These phenomena are
explored through systematic numerical simulations. Our study reveals
complicated relaxation processes as well as a crossover between two very
different steady-state regimes. 
\end{abstract}

\maketitle


\DeclareGraphicsExtensions{.pdf,.png,.jpg}



\section{Introduction}

All interesting phenomena in nature arise from many interacting degrees of
freedom. While statistical mechanics, as developed by Boltzmann and Gibbs,
provides a sound basis for understanding physical systems in thermal
equilibrium, nearly all other fascinating phenomena around us are due to
non-equilibrium stochastic processes, being coupled to more than one
reservoir (of energy, particles, etc.) Examples include all living organisms
and the life-sustaining ocean-atmosphere complex. Yet, an overarching
principle for far-from-equilibrium statistical mechanics remains elusive,
despite considerable recent progress on fluctuation theorems and the
`non-equilibrium\ counterpart' of the free energy
(for a recent, comprehensive review, see, e.g., \cite{Seifert12}). In
particular, even when a system evolves according to a master equation with
time independent rates, the probability distribution of the stationary state
is far from the simple Boltzmann factor as soon as the dynamics violates
detailed balance. Further,\ although a systematic construction for this
distribution exists \cite{Hill66}, finding its analytic form explicitly is a
highly non-trivial task. In addition, there are persistent stationary
probability current loops in such states \cite{ZS07}, leading to observable
consequences in general. Needless to say, when the rates are themselves
time-dependent (e.g., diurnal/seasonal heating/cooling of our atmosphere),
the system displays more intriguing, but less comprehensible, behavior.

One way to make progress, given the general difficulties in exploring
non-equilibrium statistical mechanics, is to study simple model systems.
Their behavior can be easily simulated by computers and their simplicity may
allow us to develop a full understanding, providing some insight into
non-equilibrium processes. In this spirit, many studies were conducted for
the paradigmatic Ising model with nearest neighbor interactions, driven to
non-equilibrium steady states (NESS) by a variety of mechanisms. In all
cases, the key lies in coupling the system to two (or more) energy
reservoirs, such as a bath and an external drive, or two thermal baths at
different temperatures. In general, there is a stationary net flux of energy 
\textit{through} our system, from say, the hotter bath to the cooler one, as
the system settles into a NESS. Over the last three decades, a wealth of
surprising phenomena associated with such NESS have been discovered, many of which
remain poorly understood. One general rule gleaned so far is that, while
driving a system into a NESS will produce novel effects in general, the most
dramatic differences tend to emerge when the system is endowed with a 
\textit{conservation }law (e.g., particle conservation). For example, taking
a model with spin-flip dynamics out of equilibrium produces
observable effects, but the critical properties, say, remain in the
equilibrium universality class \cite{Grin85,BasSch94,Trim06,TD14}. This
study is devoted to an Ising system with a conservation law in arguably the
simplest of settings: the lattice gas on a ring (one-dimensional periodic
lattice).

Before delving into our specific system, let us provide a brief overview of
the various ways in which the Ising model has been coupled to \textit{two}
thermal baths and the NESS behavior that emerged. We believe such a
paragraph will be helpful for readers encountering the term
\textquotedblleft two-temperature Ising model\textquotedblright\ in the
literature. In typical textbooks on statistical mechanics, only the static
properties of the Ising model are presented: A spin taking on two values, $%
\sigma _{i}=\pm 1$, is located on each site, $i$, of a lattice in $d$
dimensions, subjected to a variety of boundary conditions, with
ferromagnetic interactions between nearest neighbor spins. Thus, the energy
functional (Hamiltonian) associated with a configuration of spins $\left\{
\sigma _{i}\right\} $ is given by ${\mathcal H} \left[ \left\{ \sigma
_{i}\right\} \right] =-J\Sigma _{i}\sigma _{i}\sigma _{i+1}$, with $%
J>0$. Accordingly, the probability to find $\left\{ \sigma _{i}\right\} $,
when the system is in contact with a thermal bath at temperature $T$, is given by the
Boltzmann factor $P_{eq}\left[ \left\{ \sigma _{i}\right\} \right] \propto
e^{-\mathcal{H}/k_{B}T}$, while averages of observable quantities, $\mathcal{%
O}$ $\left[ \left\{ \sigma \right\} \right] $, are found by computing%
\begin{equation}
\left\langle \mathcal{O}\right\rangle \equiv \sum_{\left\{ \sigma \right\} }%
\mathcal{O}\left[ \left\{ \sigma \right\} \right] P_{eq}\left[ \left\{
\sigma \right\} \right]~.
\end{equation}%
Promoting this model to a \textit{kinetic} one, two common forms of dynamics
are used: Glauber, spin-flip \cite{Glau63}, or Kawasaki, spin exchange \cite%
{Kawa70}. In the former, a spin is chosen at random, and flipped according
to some probabilistic rule (which depends on $\mathcal{H}$ and $T$ \cite{footnote1}), so that
the total magnetization, $M\equiv \Sigma _{i}\sigma _{i}$, fluctuates in
time. By contrast, in the latter, a random nearest neighbor pair of spins
are chosen and exchanged. Thus, $M$ remains constant, a dynamics
particularly suitable for describing say, binary alloys (with spin $\pm 1$
representing, say, Cu and Zn). This version is also known as the Ising
lattice gas \cite{YangLee52}, with spin $\pm 1$ referred to as particle and
hole, a language we will mostly use here. To achieve thermal equilibrium, a
unique $T$ enters these rules no matter which spin or pair is chosen. If we
insist on coupling our system to \textit{two} $T$'s, then it is clear that
there is an enormous variety of ways to implement them. The brief survey
below provides the context of our study.

Although spin-flip rates which depend on two temperatures were introduced as
early as 1982 \cite{Hill82}, the bulk of such explorations was carried out
in the 90's (see,
e.g., \cite{SZ95} for a review of these early studies). Nearly all
studies involve dynamics which are (essentially) translationally invariant,
while many involve anisotropy. Examples include coupling to the two baths
every spin or every other spin (for Glauber dynamics in $d\geq 1$) \cite%
{TTIS} and exchanges of pairs in the $x$ or $y$ directions (in $d=2$) \cite%
{TTLG}, while more exotic models involve Glauber at one $T$ and Kawasaki at
another $T$ \cite{TTmix}. By contrast, our study will focus on \textit{%
inhomogeneous} couplings: one entire region of the system coupled to one bath and
the complement coupled to the other bath. Recent efforts were directed towards (a)
Glauber dynamics on two semi-infinite chains ($d=1$), coupled to two baths
and joined at the ends \cite{MOL} and (b) Kawasaki exchange on two halves of a
finite $d=2$ system \cite{Plei10,Li12}. While exact analytic results are
available for the former, the novelties of the NESS can be expected. For the
latter, we have only simulation results, which provided more exotic and
surprising behavior, e.g., convection cells and states with multiple strips.
Naively, the contrast between two sets of results might be attributed to the
difference in $d$: the absence/presence of a phase transition. Yet, the
other difference, Glauber \textit{vs.} Kawasaki dynamics, may be more
crucial. It is in this context that we conduct the present study: an Ising
chain with spin-exchange dynamics, coupled to two $T$'s.

The rest of this paper is organized as follows. In the next section, we
present a detailed description of the model. Although the static, equilibrium
properties of the standard Ising chain can be found in most textbooks, two
aspects $-$ Kawasaki dynamics and fixed $M$ ensemble $-$ are less well known
and will be discussed in Section III. 
We also present some results for an equilibrium model where different
coupling constants are used in different sectors of the system.
In the following Section, we report
the many surprising phenomena discovered through simulations
of the two-temperature model and compare these results with those
obtained for the standard equilibrium Ising model as well as for an
equilibrium two-coupling model. We conclude
with a summary. Some technical details are provided in Appendices.

\section{Model description}

We consider the standard Ising system on a ring of spins, $\sigma _{i}$, at $%
i=1,...,L$ sites with nearest neighbor, ferromagnetic interactions. Here, we
provide a detailed description of how we couple this system to two thermal
baths, in terms of what is implemented in our simulations.

First, we evolve a configuration by Kawasaki exchange with Metropolis rates 
\cite{Metr53}: At each time step (attempt), choose a random pair of nearest
neighbor spins and exchange them with the following probability. Clearly, a
non-trivial update must involve two spins that are opposite. If $\Delta 
\mathcal{H}$ denotes the change in $\mathcal{H}$ due to the exchange, then
we allow it to take place with probability $\min [1,e^{-\Delta \mathcal{H}%
/k_{B}T}$], where $T$ is the temperature of the thermal bath. Thus,
the overall magnetization of the system, $M$, is conserved. Since there
is no phase transition in the standard Ising model, it is natural for us to
restrict ourselves to systems with $M=0$. Of course, these systems will settle
into an equilibrium state associated with the $M=0$ ensemble.

Our goal is to explore NESS, associated with a dynamics which violates
detailed balance. One natural way is to couple two sectors of the ring to
reservoirs of differing temperatures. Specifically, exchanges within a
`window' of length $w$ will be updated with temperature $T_{w}$, while the
rest of the ring will be coupled as above. Obviously, the system will revert
back to an equilibrium Ising model for $T_{w}=T$ or $w=0$. To simplify our
study, instead of exploring the full control space of $T$-$T_{w}$, we set $%
T_{w}=\infty $ in the following. With Metropolis rates, this choice implies
that all attempts at exchanging pairs within the window are successful,
regardless of $\Delta \mathcal{H}$. One question naturally arises: How do
such exchanges differ from the case involving $J=0$, i.e., non-interacting
spins? We will address this subtle issue at the end of this section, along
with a crucial discussion of detailed balance violation.

\begin{figure}[h]
\centering
\framebox[1.3\width]{\includegraphics[scale=0.75]{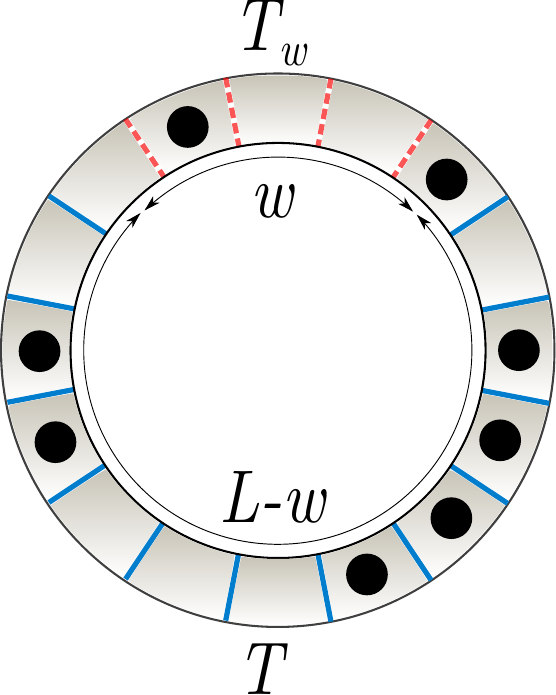}} 
\caption{(Color online) Sketch of the two-temperature Ising ring model studied in this
paper. Whereas exchanges across solid (blue online) borders are
accepted with the ordinary Metropolis rate for a system at temperature $%
T $, exchanges across dashed (red online) borders are accepted with the corresponding
rate for a system at temperature $T_{w}$.}
\label{figure1}
\end{figure}

How these rates operate is concisely captured in Fig. \ref{figure1}, where the
lattice is depicted as being half-filled with particles/holes (corresponding
to $M=0$). Any exchange takes place across a `border' between two
adjacent sites. Of the $L$ borders, we color $w$ of them red and the rest
blue. An exchange across a dashed red border always takes place. By contrast, to
exchange a pair across a blue border, $\Delta \mathcal{H}$ must be computed.
Then the exchange is allowed with probability $\min [1,e^{-\Delta \mathcal{H}%
/T}]$ (i.e., the rate used to study the ordinary equilibrium Ising model at $%
T$). \ In the following, we will refer to this as the $2T$ model, short for
`two-temperature Ising model.' Let us emphasize that all properties of this
system are embodied in a time dependent distribution, $P\left[ \left\{ \sigma
\right\} ;t\right] $, governed by a simple master equation:%
\begin{equation}
P\left[ \left\{ \sigma ^{\prime }\right\} ;t+1\right] =\sum_{\left\{ \sigma
\right\} }W\left[ \left\{ \sigma ^{\prime }\right\} ,\left\{ \sigma \right\} %
\right] P\left[ \left\{ \sigma \right\} ;t\right]   \label{ME}
\end{equation}%
where $W$ is the transition probability from $\left\{ \sigma \right\} $ to $%
\left\{ \sigma ^{\prime }\right\} $ in one attempt:%
\begin{eqnarray}
L^{-1}\sum_{i}\Delta _{i}\left[ 
\left\{ 1-q_{i}\delta \left( \sigma _{i-1}-\sigma _{i}\right) \delta \left(
\sigma _{i+1}-\sigma _{i+2}\right) \right\} \delta \left( \sigma
_{i}^{\prime }+\sigma _{i}\right) \delta \left( \sigma _{i+1}^{\prime
}+\sigma _{i+1}\right) \right.  \nonumber \\ 
\left. +q_{i}\delta \left( \sigma _{i-1}-\sigma _{i}\right) \delta \left( \sigma
_{i+1}-\sigma _{i+2}\right) \delta \left( \sigma _{i}^{\prime }-\sigma
_{i}\right) \delta \left( \sigma _{i+1}^{\prime }-\sigma _{i+1}\right) 
\right]  \label{W}
\end{eqnarray}%
Here, $\delta $ is the Kronecker delta
(i.e., unity if its argument vanishes and zero otherwise), 
$\Delta _{i}\equiv \delta \left( \sigma _{i} + \sigma _{i+1} \right)
\Pi _{k\neq i,i+1}\delta \left( \sigma _{k}^{\prime }-\sigma _{k}\right) $ 
ensures that
only the pair $\sigma _{i},\sigma _{i+1}$ may change, and%
\begin{equation}
q_{i}=1-e^{-4J/k_{B}T(i)}  \label{2T-q}
\end{equation}%
with $T(i)=T_{w}$ for $i=1,...,w$ and $T(i)=T$ for $i=w+1,...,L$, is the
probability that this pair is unchanged. Note that $\sum_{\left\{
\sigma ^{\prime }\right\} }W\left[ \left\{ \sigma ^{\prime }\right\}
,\left\{ \sigma \right\} \right] =1$ so that $P$ remains normalized at all
times. It can be shown that, as $t\rightarrow \infty $, $P$ settles into a
unique stationary distribution, $P^{\ast }\left[ \left\{ \sigma \right\} %
\right] $, which is very different from the Boltzmann $P_{eq}$ (with any $T$%
). Indeed, for small systems ($L=6,8$), we found explicit distributions to
have very different degeneracy structures than the equilibrium distributions for the same
$\mathcal{H}$. For reasonably large systems (e.g., $L\gtrsim 50$), finding
these $P^{\ast }$'s is impractical numerically and analytically (due to
detailed balance violation), let alone computing averages of observables
from $P^{\ast }$. Thus, Monte Carlo simulations are the only viable technique
for us to make progress \cite{footnote}.

When we consider the exchanges away from the interface between the two
sectors, the rates resemble those for an inhomogeneous Ising model (with $%
J=0 $ inside the window) coupled to a single reservoir at temperature $T$.
Specifically, we can still exploit Fig. \ref{figure1}, by regarding the blue
borders as energy bonds of strength $J$ and associating the red ones with $%
J=0$. Let us refer to this model as the $2J$ model. Clearly, the stationary
distribution here is just $\exp \left[ -J\Sigma _{i}\sigma _{i}\sigma
_{i+1}/k_{B}T\right] $, where the sum is over only the blue bonds. Its
statistical properties are just as accessible as the standard model. What is
the key difference between the transition rates for this model and the $2T$
case? It lies in the exchanges of just two pairs of spins at each interface.
To see this mathematically, we note that the dynamics needed for a generally
inhomogeneous Ising model (i.e., ${\mathcal H} =-\Sigma _{i}J(i)\sigma
_{i}\sigma _{i+1}$) in contact with a single bath are the same as above,
except for 
\begin{equation}
q_{i}=1-e^{-2 \left( J(i-1)+J(i+1)\right) /k_{B}T}  \label{2J-q}
\end{equation}%
instead of Eq. (\ref{2T-q}). For the $2J$ model, we have $J(i)=0$ and $J$,
for $i\in \left[ 1,w\right] $ and $\left[ w+1,L\right] $ respectively. How
does one set of rates obey detailed balance and the other set violate it?
The answer can be found in Appendix A.

To end this section, let us provide the details of our simulation methods.
Starting with a random configuration of spins with $M=0$ if $L$ is even ($M=1
$ for odd $L$), we randomly choose a pair of spins and update them according
to the probabilities in Eq. (\ref{ME}). A Monte Carlo step (MCS) is defined
as $L$ such attempts. Whereas our primary concern is with the steady-state
properties, we also need to ensure that the system has relaxed sufficiently
into the NESS. For this purpose, we also collected data on the transient
regime. After the relaxation stage, a suitable number of simulation steps is
used to measure the averages of various observables. Specifically, we will
focus on two-spin correlations 
\begin{equation}
G\left( i;r\right) \equiv \left\langle \sigma _{i}\sigma _{i+r}\right\rangle 
\label{Gi}
\end{equation}
(for certain $i$'s), the total magnetization within the window%
\begin{equation}
M_{w}\equiv \sum\limits_{i=1}^{w}\sigma _{i}~~,  \label{Mw}
\end{equation}
the distribution of the normalized window magnetization $m\equiv M_{w}/w$%
\begin{equation}
\mathcal{P}_{w}\left( m\right) \equiv \sum_{\left\{ \sigma \right\} }\delta
\left( m-\sum\limits_{i=1}^{w}\sigma _{i}/w\right) P^{\ast }\left[ \left\{
\sigma \right\} \right] ~~,  \label{Pw}
\end{equation}
the energy density profile 
\begin{equation}
u_{i}\equiv -J\left\langle \sigma _{i}\sigma _{i+1}\right\rangle =- J G\left(
i;1\right)   \label{u}
\end{equation}%
and its sum $\Sigma _{i=1}^{L}u_{i}$ (i.e., $\left\langle \mathcal{H}%
\right\rangle $). The number of relaxation steps
required varies considerably with the system parameters and
increases rapidly for larger $L$'s and smaller $T$'s, see the discussion below.
Finally, we performed typically 40 to 100
independent runs (with different initial conditions), so that time dependent
quantities are constructed by averaging over these runs. Of course, for
stationary state properties, we perform both a time and ensemble average.
In the remainder of this paper, we
choose units such that $J/k_{B}=1$.

\section{The equilibrium Ising lattice gas}

\begin{figure}[h]
\centering
\includegraphics[scale=.80]{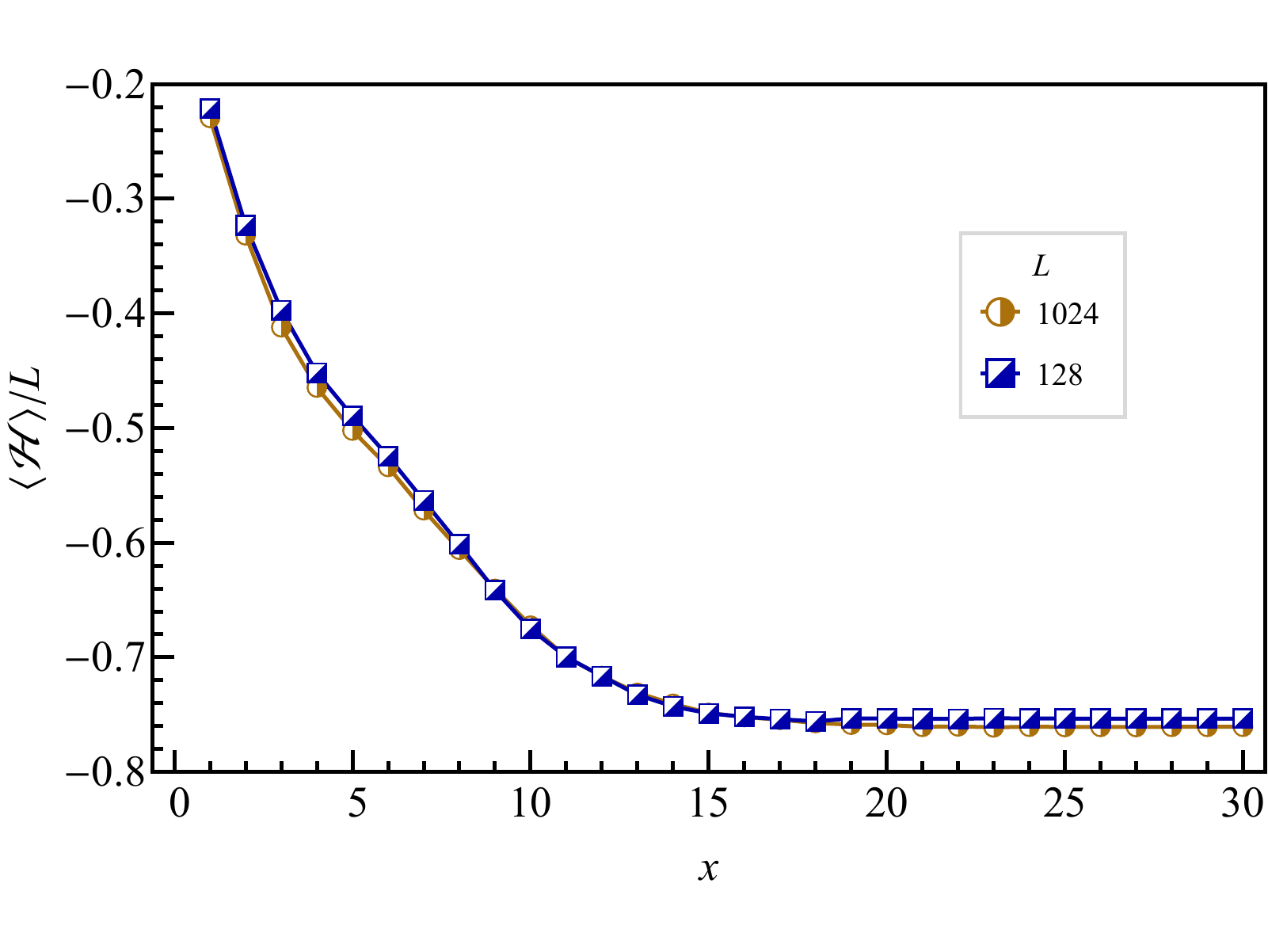}
\caption{(Color online) Log-binning relaxation of the energy density $\langle {\mathcal H} \rangle/L$
for the standard Ising system
at temperature $T=1$ and two different lattice sizes $L$. 
In this plot (as well as in the other figures below showing log-binning relaxation)
the average of the quantity of interest is sampled at intervals between $2^x$ and $2^{x+1}$ MCS.
The equilibrium value of $\langle {\mathcal H} \rangle/L$
is already very close to the value $- \tanh(1) \approx -0.76$ of the infinite lattice.
The data result from averaging over an ensemble of 40 to 100
independent realizations. Error bars are comparable to the sizes of the
symbols.}
\label{figure2}
\end{figure}

Before reporting results on the $2T$ Ising ring, we will briefly review the
relevant properties of two equilibrium cases, namely the standard homogeneous
Ising model with uniform couplings as well as the $2J$ model discussed in
the previous section.

Though the standard Ising chain
is a textbook model, we present some less well-known aspects, so as to
facilitate the comparison with its non-equilibrium counterpart. Deferring
technical details to Appendix B, we report only simulation results here.

\begin{figure}[h]
\centering
\includegraphics[scale=.80]{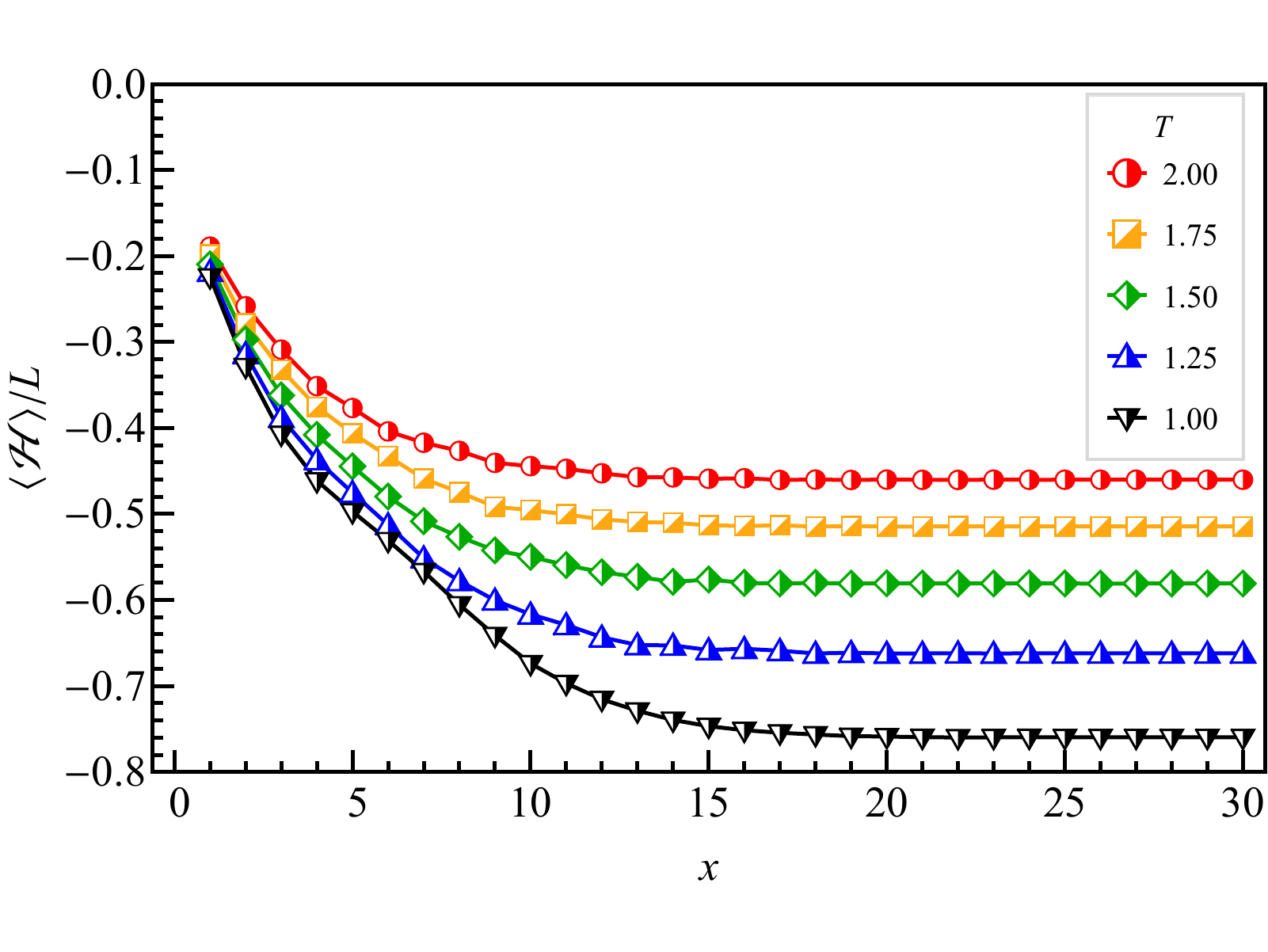}
\caption{(Color online) Log-binning relaxation of the energy density for the standard Ising model
with $L=512$. The data result from averaging over an ensemble of 40 to 100
independent realizations. Error bars are comparable to the sizes of the
symbols.}
\label{figure3}
\end{figure}

As the system relaxes into the stationary state, its behavior is dominated
by eigenvectors which are associated with eigenvalues of $W$, see Eq. (\ref{W}), close to unity.
While the details depend on the specifics of the rates, it is known that a
power law ($t^{1/3}$ law) can be expected with Kawasaki dynamics. In particular,
at low temperatures domains (of the same spin) should form and grow in
size. Of course, unlike typical coarsening behavior of a system with
spontaneous magnetization (i.e., below critical temperature), domains in the
ring cannot be much larger than the correlation length: $O\left(
e^{1/T}\right) $. Nevertheless, in the growing regime, their sizes scale
with $t^{1/3}$ \cite{Corn91}. Though we did not measure domain sizes, the
relaxation of all the quantities we study are consistent with this law.
Fig.~\ref{figure2} shows how the 
energy density $\left\langle \mathcal{H}\right\rangle /L $\ relaxes
for different $L$'s. 
We thereby use the log-binning procedure, in which the averages of quantities of interest 
are sampled at intervals between $2^x$ and $2^{x+1}$ MCS.
In Fig. \ref{figure3}, we show the effects of different $T$ on
the relaxation of $\left\langle \mathcal{H}\right\rangle /L $. Not surprisingly,
the energy of a system coupled to lower temperatures takes considerably
longer to settle. For our $2T$ model, nothing was known about such
relaxation times. Below, we will use a similar approach to determine if
those systems have settled into their NESS.

\begin{figure}[h]
\centering
\framebox[1.05\width]{\includegraphics[width=0.30%
\columnwidth]{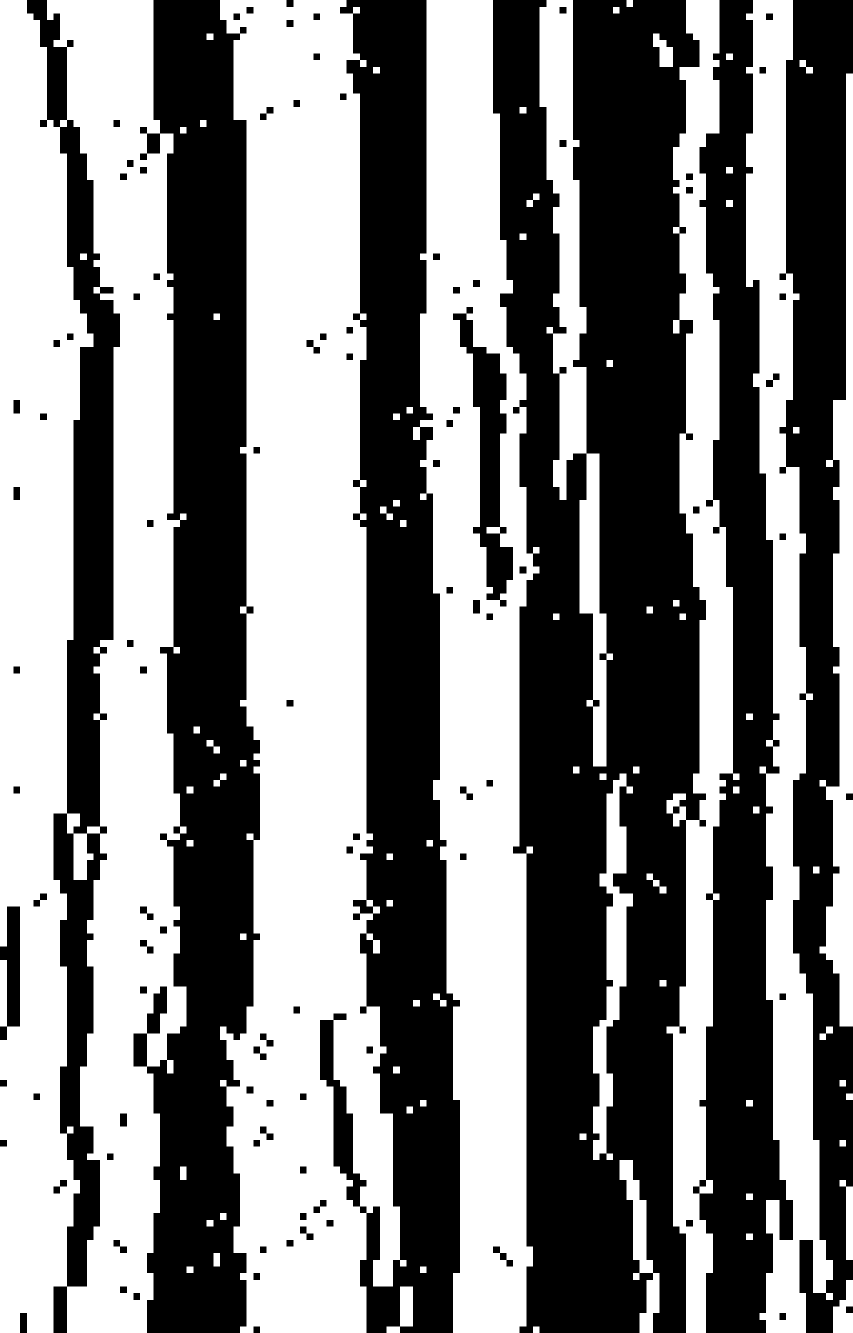}}
\caption{Time trace of the standard Ising system after it reached
equilibrium at $T=1$. The plot shows the evolution of the system over a total of 2000
MCS after an initial relaxation period of $2^{20}$ MCS, where two consecutive horizontal lines are separated by 10 MCS. The
system size is $L=128$.}
\label{figure4}
\end{figure}

Turning to properties in the stationary state, let us first illustrate a
typical time trace of the configurations, in Fig. \ref{figure4}, with a system
with $L=128$ set at $T=1$. A black/white square denotes a particle/hole.
Each row is a snap shot of the ring, while successive rows are separated by
10 MCS. The presence of semi-persistent domains is evident. Of course, they
shrink and grow randomly, via evaporation and condensation at the edges,
resulting in, at times, apparent drifts like a random walker. The figure
should provide an intuitive picture for the quantitative aspects, such as
the two-spin correlations $G\left( r \right) \equiv G\left( i;r\right) $ (for a particular $i$, but
independent of $i$ due to translational invariance) and distributions $%
\mathcal{P}_{w}\left( m\right) $. The former is a standard quantity of
interest in the study of statistical mechanics. By contrast, we are not
aware of systematic investigations of the latter. Here, $m$ is a local
magnetization, coarse-grained at the level of $w$, so that $\mathcal{P}%
_{w}\left( m\right) $ provides information on clustering at this length
scale. Of course, at the level of the entire system, $\mathcal{P}\left(
M\right) $ (for the non-conserved case) enjoyed considerable attention,
since it signals the onset of long range order and reveals non-trivial
critical properties (for Ising models in $d>1$). Since we will focus on $%
\mathcal{P}_{w}\left( m\right) $ for the $2T$ model, collecting data on its
equilibrium counterpart will provide both a good baseline and a sharp
contrast.

\begin{figure}[h]
\centering
\includegraphics[scale=.80]{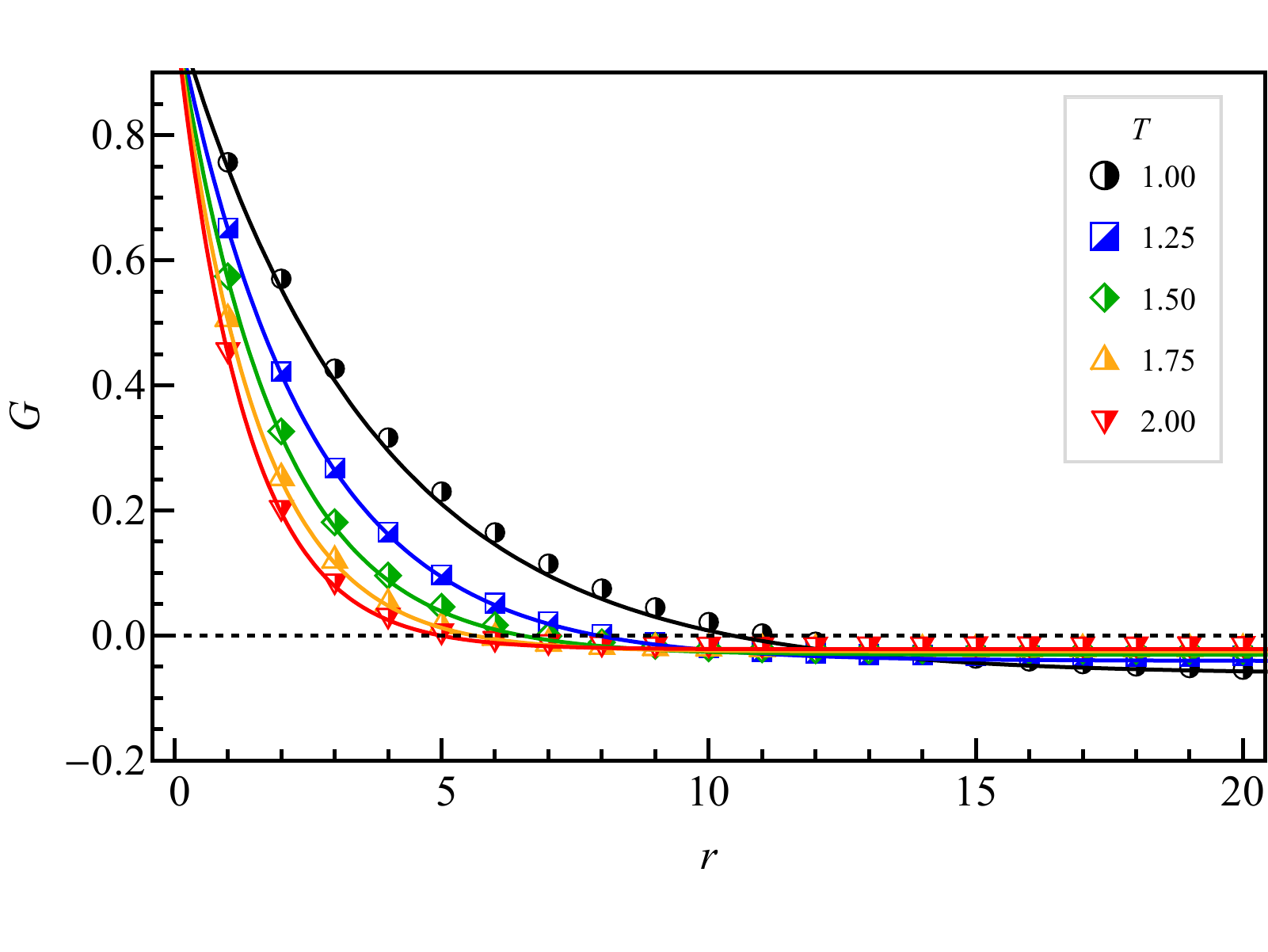} 
\caption{(Color online) $G(r)$
for the standard equilibrium Ising ring with Kawasaki dynamics for a system of length 
$L=127$. L is chosen to be odd to ensure a symmetric result about a given reference
point. The lines refer to the approximation Eq. (\protect\ref{eq:G_linear}).}
\label{figure5}
\end{figure}

\begin{figure}[h]
\centering
\includegraphics[scale=.80]{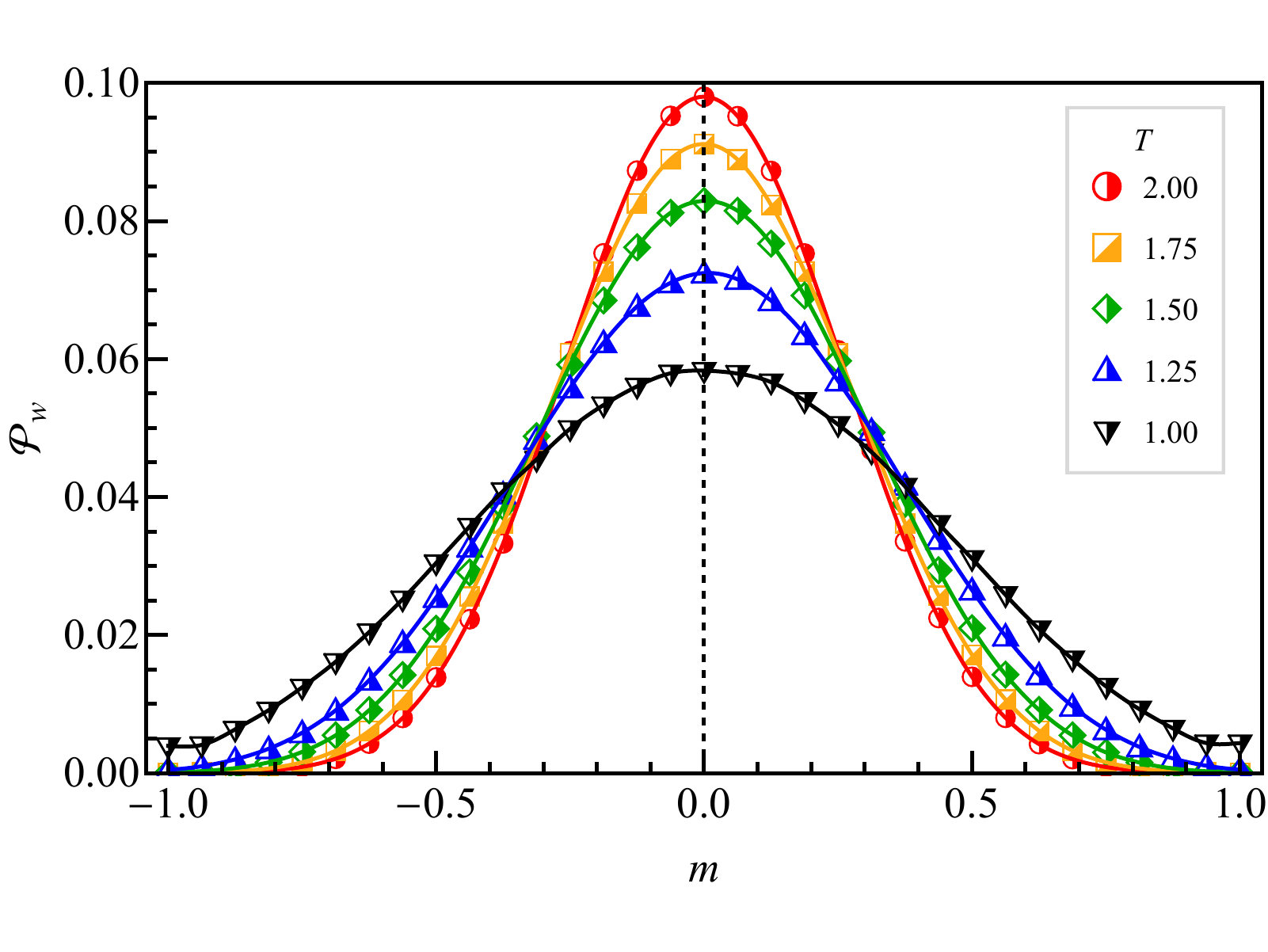} 
\caption{(Color online) Equilibrium distributions of the normalized window magnetization $m$
for the standard Ising model
with $L=128$ and $w=32$.}
\label{figure6}
\end{figure}

For a non-conserved Ising model on a ring, $G$ is well-established: 
\begin{equation}
G(r) = \left(
\omega ^{r}+\omega ^{L-r}\right) /\left( 1+\omega ^{L}\right) , 
\end{equation}
where 
\begin{equation}
\omega \equiv \tanh \left( 1/T\right) ~.
\end{equation}%
Of course, $G>0$ for all finite $T$. But the $M=0$ constraint adds
complications, since $\Sigma _{r}G\left( i;r\right) =0$ in this case. With
details deferred to Appendix B, we find that a simple linear approximation 
\begin{equation} \label{eq:G_linear}
G\cong A\left( \omega ^{r}+\omega ^{L-r}\right) +B
\end{equation}%
(with $A$ and $B$ fixed by $G(0)=1$ and $\Sigma G=0$)
agrees quite well with data. In Fig.\ \ref{figure5}, we illustrate this
agreement in a small system with $L=127$ sites. There are no surprises here; its
main purpose is for comparison with Fig. \ref{figure13} below.

Turning to $\mathcal{P}_{w}\left( m\right) $, we illustrate with Fig. \ref%
{figure6} the case of $L=128$ and $w=32$. For each $T$, we compile a histogram
from $\thicksim 10^{8}$ measurements of $M_{w}$. 
Not surprisingly, every
distribution is peaked at $m=0$, the signature of disorder. In stark
contrast, below we will find transitions to bi-modal distributions in the $2T
$ model, shown in Fig. \ref{figure15}.

\begin{figure}[h]
\centering
\framebox[1.05\width]{\includegraphics[width=0.30%
\columnwidth]{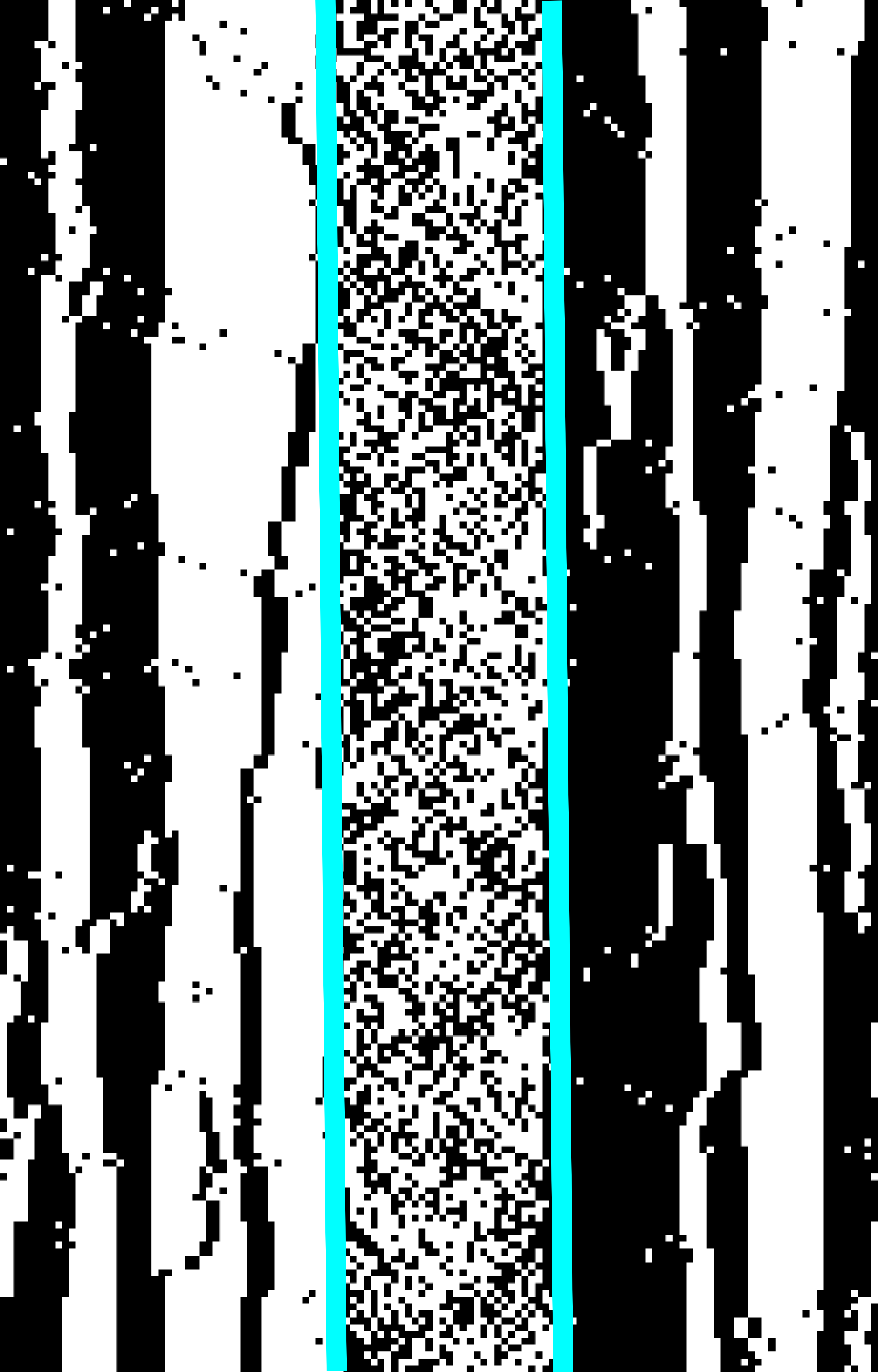}}
\caption{(Color online) Time trace of the $2J$ model after it reached
equilibrium at $T=1$. The plot shows the evolution of the system over a total of 2000
MCS after an initial relaxation period of $2^{20}$ MCS, where two consecutive horizontal lines are separated by 10 MCS. The
system size is $L=128$. The boundaries between the regions are highlighted in cyan (grey).}
\label{figure7}
\end{figure}

\begin{figure}[!h]
\centering
\includegraphics[scale=.80]{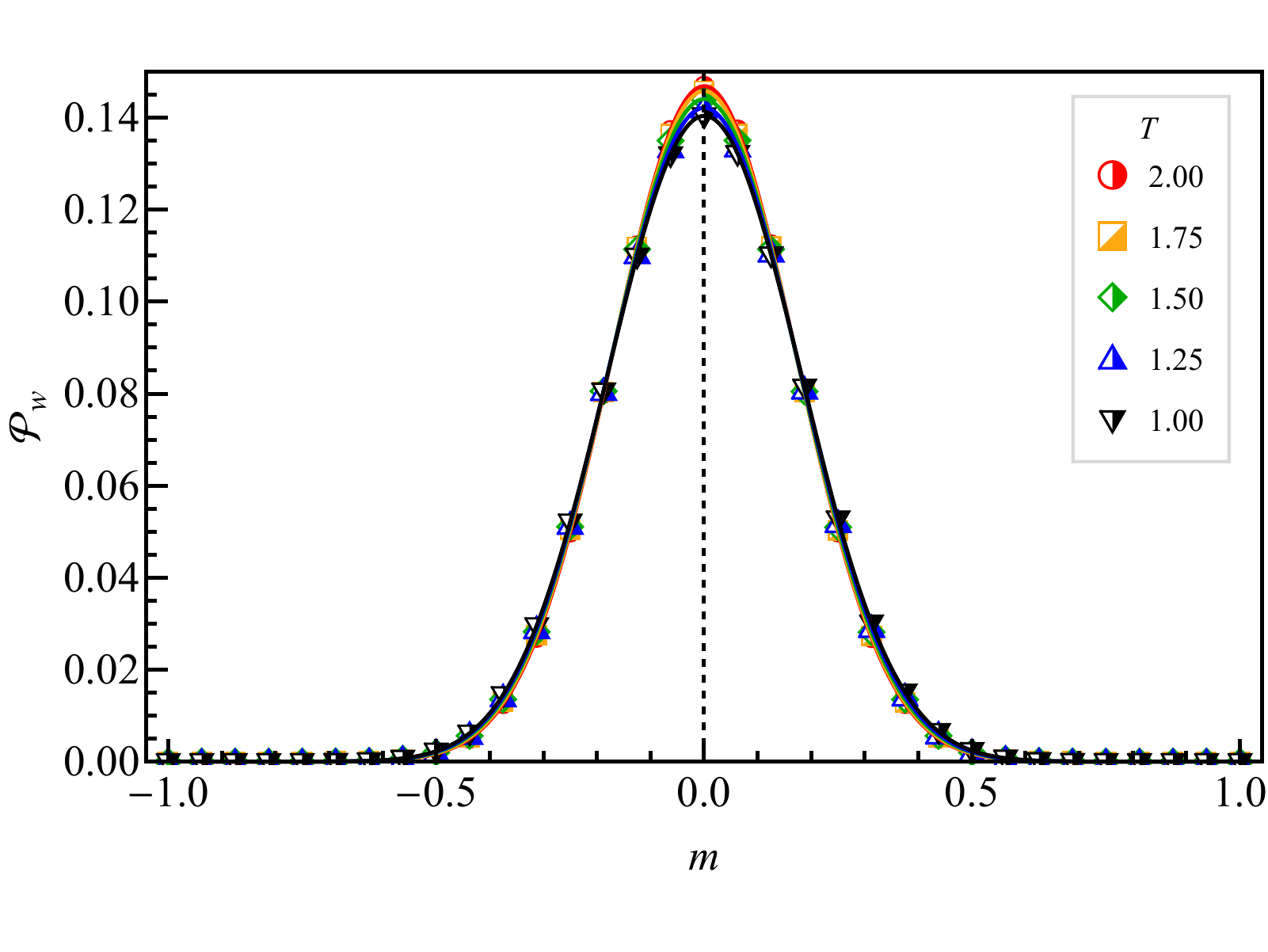}
\caption{(Color online) Equilibrium distributions of the normalized window magnetization $m$
for the $2J$ Ising model
with $L=128$ and $w=32$. Distributions for different temperatures are shown.}
\label{figure8}
\end{figure}

Before moving on to the $2T$ model let us briefly discuss some aspects of the equilibrium $2J$ model where
inside a window of width $w$ the coupling constant is zero, whereas outside of that window it is $J$. As 
discussed in the previous Section as well as in Appendix A, the sole difference between the $2J$ and the $2T$
models are the transition rates for exchanges of pairs of spins at the interface. For that reason, comparison
between results obtained from the $2J$ and $2T$ models will be very enlightening.

Fig.~\ref{figure7} shows the time trace for the $2J$ model after having reached equilibrium. One distinguishes
two regions, namely the region inside the window which is disordered, due to the absence of couplings between spins
in that region, and the region outside with semi-persistent domains, similar to what is observed in the standard
Ising model, see Fig.~\ref{figure4}. As the presence of the interface does not create long-range correlations, 
the behavior of the system outside of the window is largely unaffected by the presence of the disordered sector.

Equilibrium distributions of the normalized window magnetization are displayed in Fig.~\ref{figure8}
for the $2J$ model. As expected
for a disordered section, these distributions are Gaussian and only show a very weak dependence on the
temperature of the system. 

\section{The two-temperature non-equilibrium model}
We now proceed to discuss the two-temperature non-equilibrium ring 
and the intriguing and unexpected features that emerge from a local 
breaking of detailed balance. The investigation of steady-state
and relaxation properties is mainly done through the same quantities as those introduced in
the previous Section for the characterization of the standard Ising model.

\subsection{Steady-state time traces}

A good starting point for appreciating the nature of the non-equilibrium
steady states that arise in the two-temperature ring are the time traces of
systems that have settled into the steady state.

\begin{figure}[h!]
\centering
\framebox[1.05\width]{\includegraphics[width=0.30%
\columnwidth]{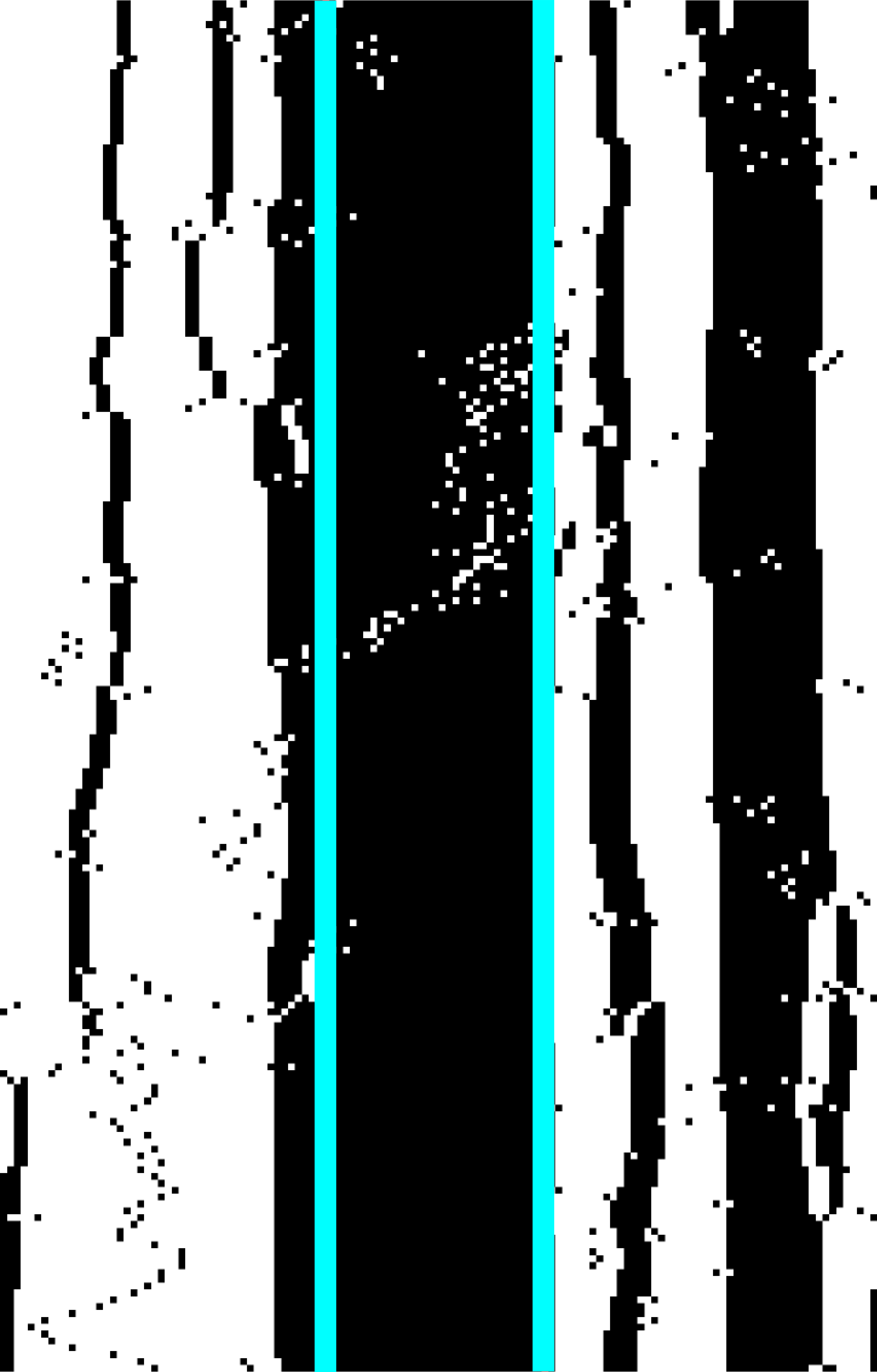}} ~~~~~~~~~ \framebox[1.05\width]{%
\includegraphics[width=0.30\columnwidth]{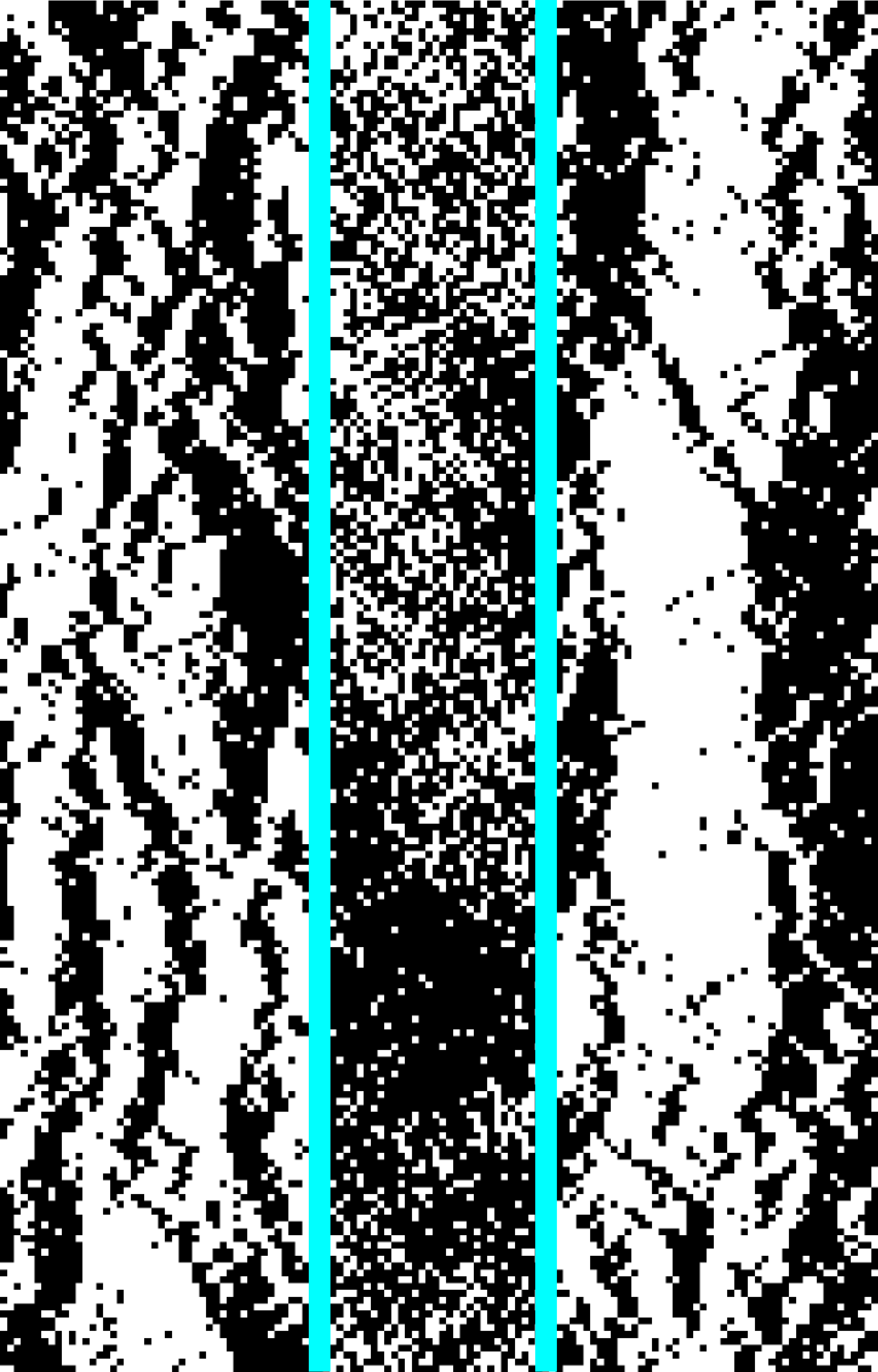}}
\caption{(Color online) Time traces of the two-temperature model at different values of $%
T $. The traces are from a system of length $L = 128$, with a window of size $w
= 32$ wherein the system is in contact with a reservoir at temperature $%
T_w=\infty$. For the left system we have $T = 1$, 
whereas for the right system $T = 1.75$. The
boundaries between the regions are highlighted in cyan (grey).
Both plots show the evolution of the system over a total of 2000
MCS after an initial relaxation period of $2^{20}$ MCS, where two consecutive vertical lines are separated by 10 MCS.}
\label{figure9}
\end{figure}

The time trace shown in the left panel of Fig.\ \ref{figure9}, with the
temperature outside the window being at $T=1$, is in stark contrast to the
time traces in Figs.\ \ref{figure4} and \ref{figure7} for the standard Ising
system and the equilibrium $2J$ system at the same
temperature. While phase-separated domains still form as in the equilibrium
model, the presence of one large domain which envelopes the central window
(indicated by cyan lines) is noteworthy. This domain 
remains pinned to the location of the infinite temperature window. 
This behavior is particularly strange when
considered in light of the equilibrium system at
infinite temperature where the spins are
distributed in a completely uncorrelated manner. It is thus
surprising that the infinite temperature window of the two-temperature
model is the most ordered region of the lattice!

The time traces provide
strong indications of a crossover between different steady-state
regimes. This transition can be observed when varying either the size of the
infinite temperature window or the temperature of the lattice sites 
\emph{outside} the window, $T$. Fig. \ref{figure9} demonstrates the effect due to
varying $T$.
At larger values of $T$ (illustrated in the right panel in Fig. \ref{figure9}),
the infinite temperature window behaves more or
less as expected by being more disordered than the much colder, surrounding
lattice. When $T$ is lowered, a crossover takes place to a state 
where the system within the window is much more ordered,
despite being coupled to a much higher temperature bath than that in the surroundings  
(illustrated in the left panel of Fig. \ref{figure9}). 

There is another perspective to this remarkable crossover, namely, when $T_{w}$ 
is raised from $T$ (for the case $T=1$, say) to $T_w=\infty$. 
Then, within the window and contrary to expectations, order will gradually 
emerge from disorder! Such surprising phenomena have been reported  
decades ago in the context of the two-temperature Ising lattice gas 
in two dimensions \cite{TTLG,SZ95}. Finding this unexpected feature -- 
the increase of order despite an apparent increase of fluctuations -- 
in driven {\it continuum} systems, Helbing, et. al. coined it `freezing by heating' 
\cite{Helb00}. In a more general setting, we may regard these counter-intuitive 
behavior as `negative response,' a property that can be expected in a wide class 
of non-equilibrium statistical systems \cite{Zia02}. 

\subsection{Relaxation process}

Before exploring and quantifying the steady-state features further, we need
to consider the issue of how the system relaxes towards the steady
state.
As for the standard equilibrium system in the previous Section we use the log-binning
procedure in order to probe the relaxation process. As shown in the following,
in addition to providing a means to
confirm entrance into the steady state, these measurements contain
important hints to the process which causes the window to display this high level of order.

\begin{figure}[h!]
\centering
\includegraphics[scale=.75]{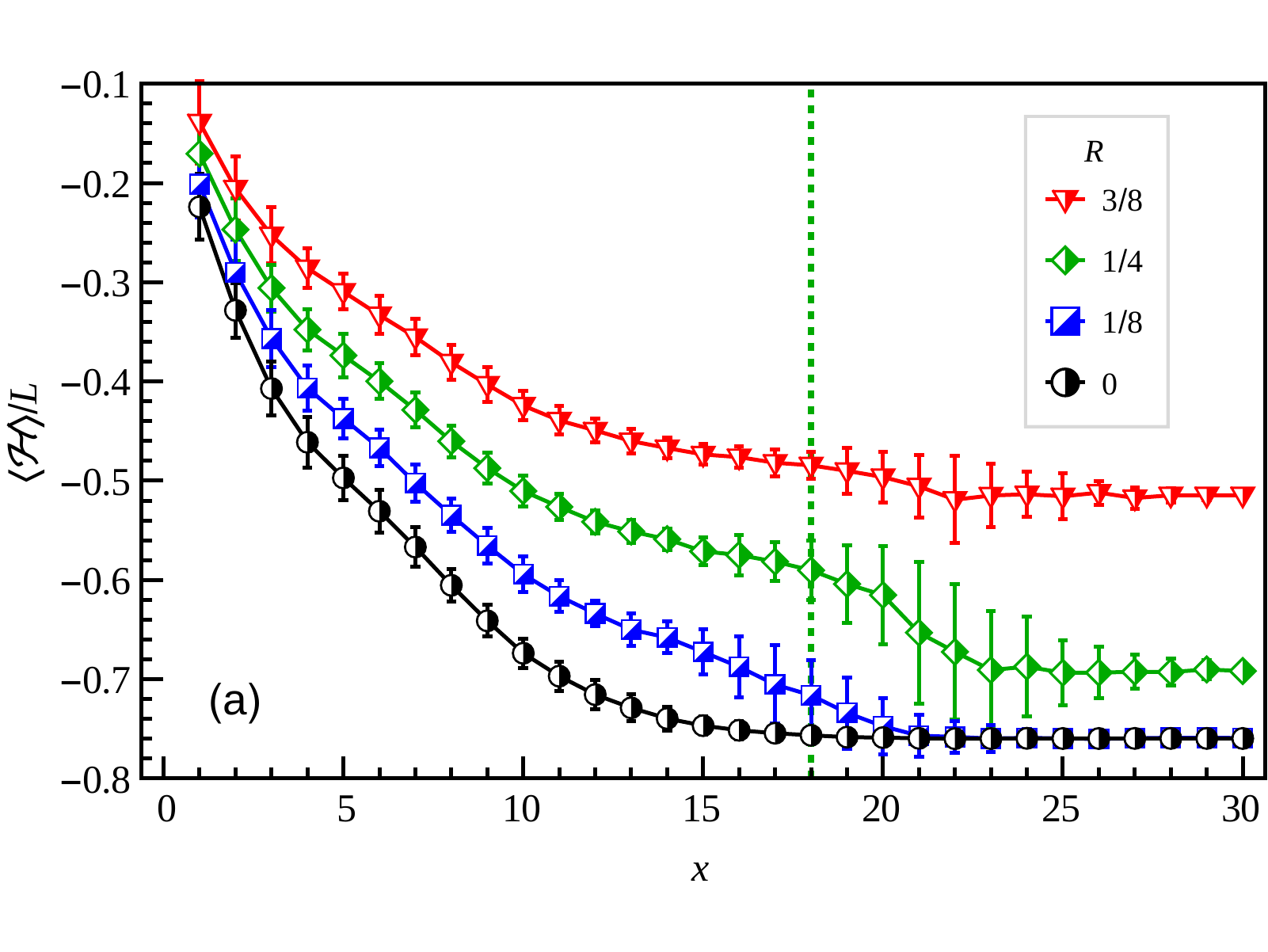}\\[-0.7cm]
\includegraphics[scale=.75]{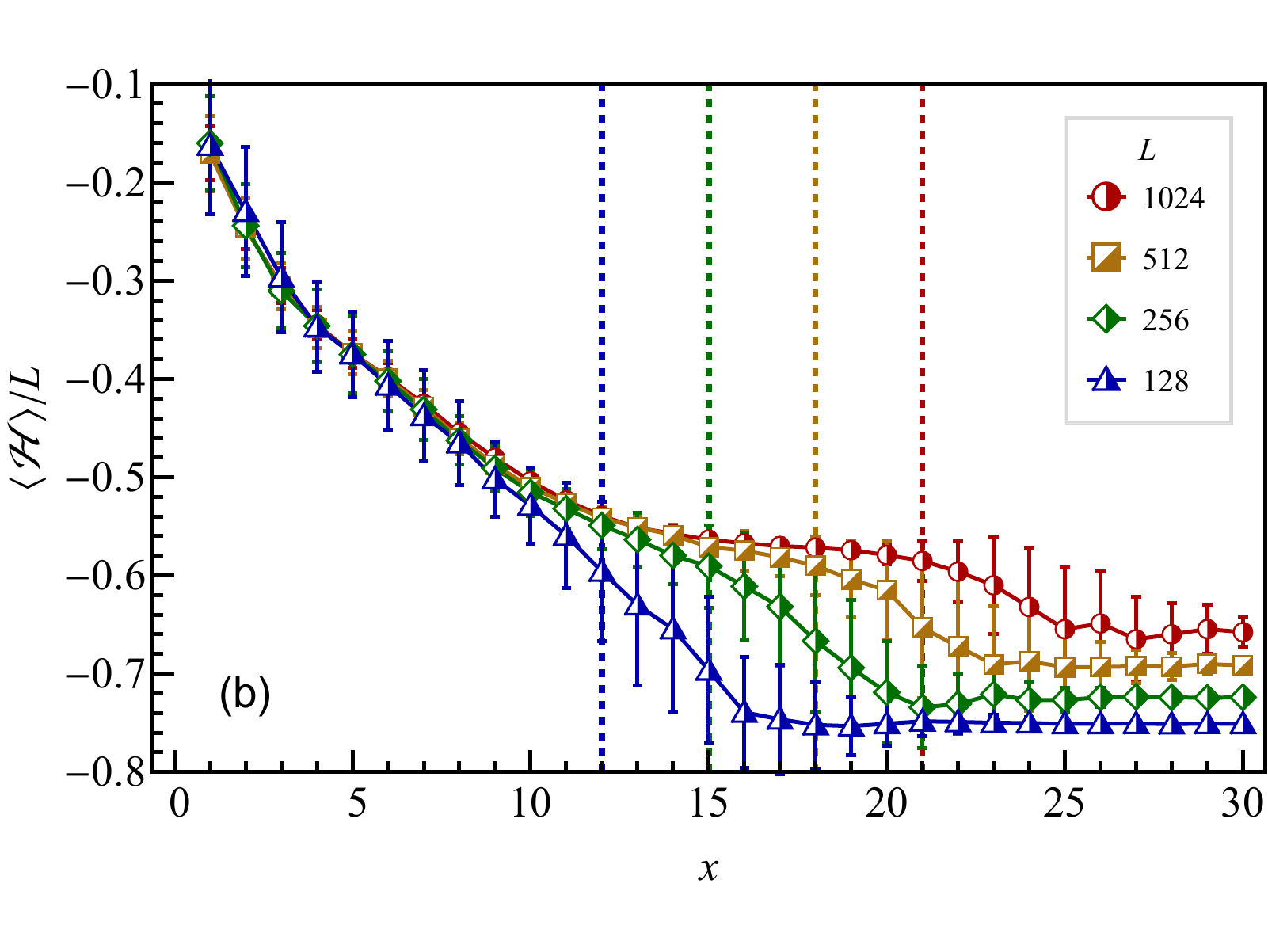}
\caption{(Color online) (a) Log-binning relaxation of the average energy density for $T =
1$ with fixed system size $L=512$ and varying window size $w$ (yielding different values
for the ratio $R = w/L$, see the figure legend). The dashed
vertical line is placed at the point at which the green curve with $w=128$ exits the
metastable state, which is equivalent to $(w/2)^3=64^3 = 2^{18}$. (b) Log-binning
relaxation of the average energy density for fixed ratio $R=w/L=
1/4$ and $T = 1$. The dashed vertical lines are placed to correspond with 
$(w/2)^3$ for their respective curves (left most curve located at $16^3$, right most curve located
at $128^3$).}
\label{figure10}
\end{figure}

The addition of an infinite temperature window within the lattice
dramatically changes the manner in which the system relaxes for smaller values of $T$, as becomes
obvious when comparing the time-dependent energy in Fig.\ \ref{figure10} with that in
Figs.\ \ref{figure2} and \ref{figure3}. Whereas Fig.\ %
\ref{figure10}a contains for $T=1$ curves with different window sizes for a
system with $L=512$ sites, Fig.\ \ref{figure10}b shows for the same temperature the relaxation of the
energy when varying $L$ and $w$ in such a way that the ratio $R=w/L$ is kept
constant.

There are a couple of particularly interesting features which distinguish
the two-temperature curves from the equilibrium ones. First, the
non-equilibrium curves are characterized by the existence of a metastable
state followed by a sudden increase in fluctuations and decline in average
energy. This metastable state is revealed by the flattening of the curve
and the reduction in statistical error which mimics the behavior of a system
entering a steady state. The duration of this metastable state is 
proportional to the size of the window, specifically scaling with $(w/2)^3$. 
This is reminiscent of the scaling seen within a one-dimensional Ising system with conserved dynamics where the
domain size scales as 
$t^{1/3}$ \cite{Corn91}, indicating that domain growth outside the window drives this phenomenon.
Using the above scaling relationship, a domain of size $w/2=64$ should take
approximately $2^{18}$ MCS to form. 
This time, indicated by the dashed
vertical line in Fig. \ref{figure10}a, roughly corresponds with the end of the
metastable state for the curve with $w=128$.

This idea is further supported by Fig. \ref{figure10}b, which displays the
average energy density for various system sizes $L$.
Here the ratio of the window to
lattice size, $R = w/L$, is kept fixed at $1/4$. After the initial slope on which
all of the curves collapse - due to initial cluster formation - the curves
then diverge as they enter the metastable state. This is to be expected if
the above reasoning holds since each curve has a different associated window
size, and thus a proportionately different time to exit the metastable
state. Again, the dashed vertical lines indicate the time this should take
for the correspondingly colored curves. As before, the match
indicates that the onset of the ordered phase scales
with $(w/2)^3$.
This points to the fact that the cluster formation outside of
the window strongly influences the emergence of the ordered window.

\begin{figure}[h!]
\centering
\includegraphics[scale=.60]{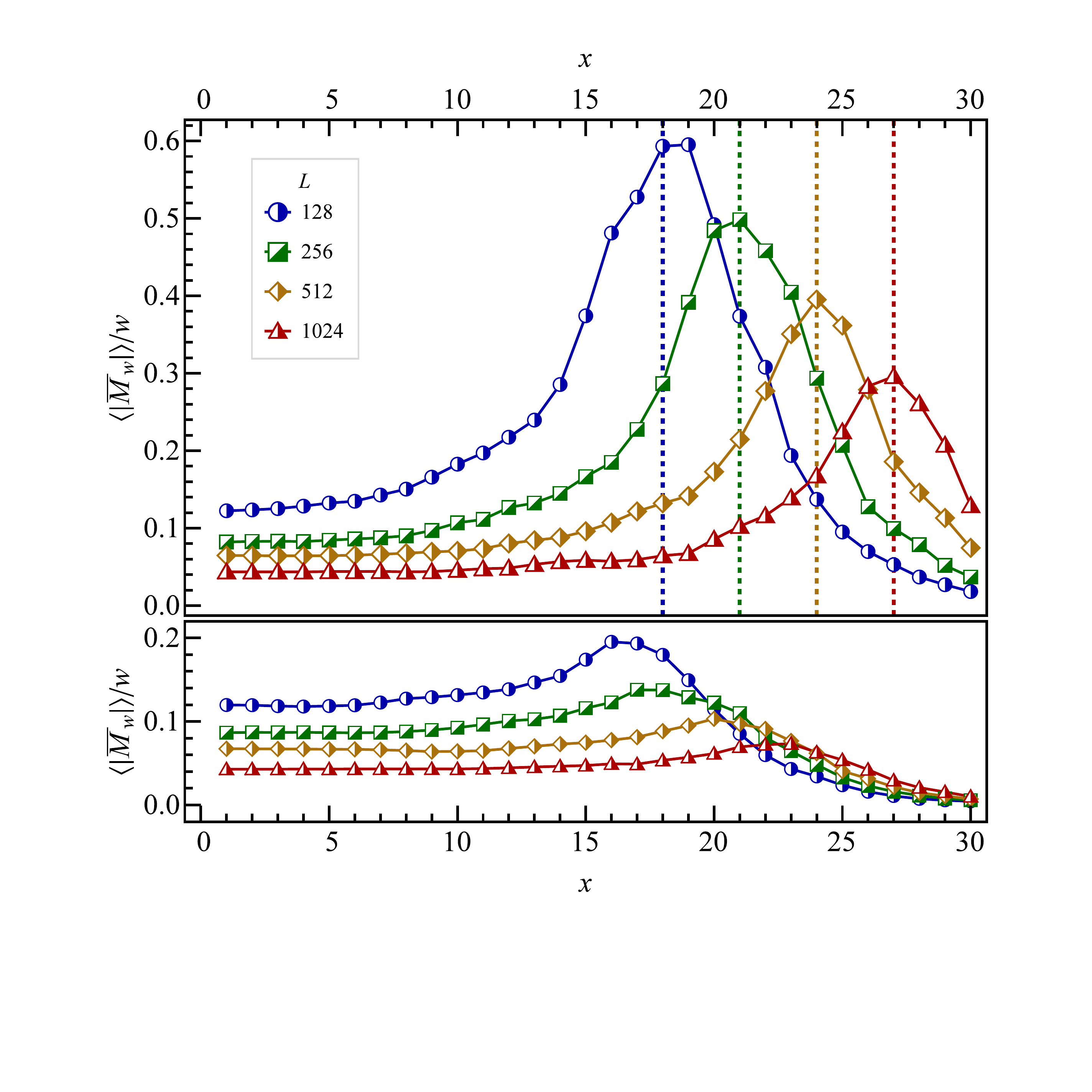}
\caption{(Color online) Upper panel: relaxation of the normalized average window
magnetization for $R = 1/4$ and $T = 1$, see main text. Here the dashed vertical lines
correspond with $(2 w)^3$. Lower panel: the same for the standard Ising model.}
\label{figure11}
\end{figure}

As a final demonstration of the importance of the $t \propto w^3$ scaling
within this system, the upper panel Fig. \ref{figure11} mirrors Fig. \ref{figure10}, but now with
the time- and ensemble-averaged window magnetization
$\left\langle |{\overline{M}_w}|
\right\rangle$ as the quantity of interest. 
We should emphasize here that for $\left\langle |{%
\overline{M}_{w}}|\right\rangle $, we first average the window magnetization
from the start to time $t$, which yields a quantity which may be either
positive or negative and eventually goes to zero as enough of phase space is
explored and the spin symmetry takes over. The absolute value of the
time-average is then taken prior to constructing the ensemble average, here
represented by the angular brackets and again being performed over 40-100
independent realizations. The construction of this quantity, while a bit
unorthodox, has a key advantage in that it not only shows distinctly how
long it takes for the ordered state to develop, but also how long it takes
for an ordered state of \emph{opposite} sign to displace the original
configuration. This point can be identified as the peak within the curves,
and the time for this to take place scales very well with $%
(2 w)^3$. In the lower panel we show for comparison the same quantity for the
standard Ising model. By construction, $\left\langle |{%
\overline{M}_{w}}|\right\rangle $ also displays  maxima for that case, but
the heights of these maxima are much smaller than for the $2T$ model. This reflects the
fact, see the time trace in Fig.~\ref{figure4}, that every choice of the window will include
both positively and negatively magnetized domains, in contrast to the non-equilibrium 
case where the window at infinity temperature is occupied by a single, almost perfectly ordered
domain.

\begin{figure}[h!]
\centering
\includegraphics[scale=.80]{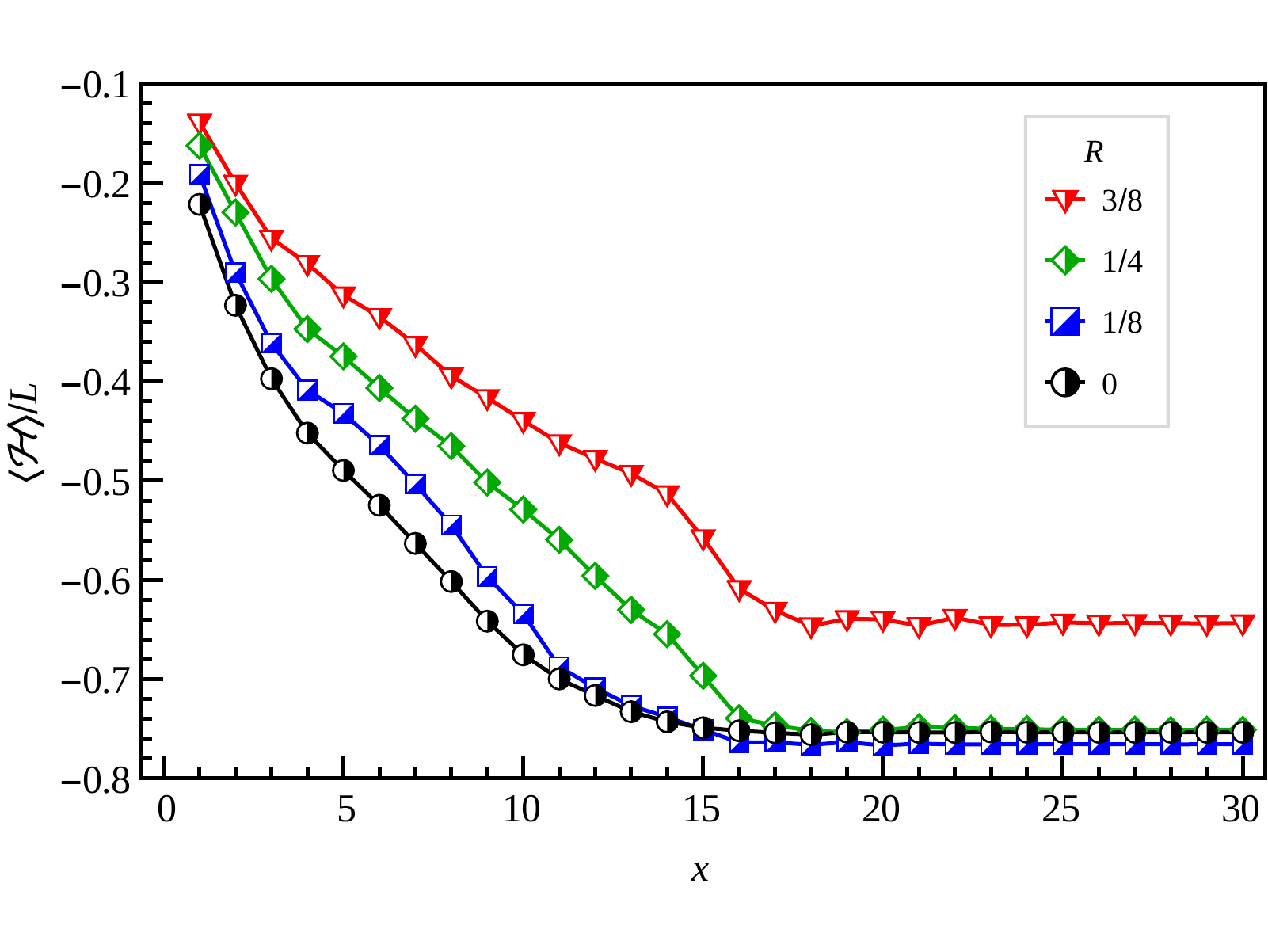}
\caption{(Color online) Log-binning relaxation of the average energy density for $L = 128$, 
$T = 1$, and various values of the ratio $R=w/L$.}
\label{figure12}
\end{figure}

Returning to Fig. \ref{figure10}a, the other point of interest is
the final relaxed level of the non-equilibrium system with 
ratio $R=1/8$ (which corresponds to window size $w =
64$). What is notable in this case is that the energy per lattice site
ultimately falls to nearly the same level as the equilibrium (black) curve.
Fig. \ref{figure12} demonstrates that it is even
possible to have \emph{lower} than equilibrium energy with the right
mix of parameters.
Notice that the curve for $R=1/8$ (which here corresponds to $w = 16$ 
as $L=128$) not only approaches the value for the
equilibrium model, but actually goes below it in the steady state. This is
admittedly achieved for a small system size in which finite-size
effects are almost certain to play a significant role.
Such a result is again reminiscent of the negative response considered in 
\cite{Zia02}.

\subsection{Steady-state properties}

\begin{figure}[h!]
\centering
\includegraphics[scale=.80]{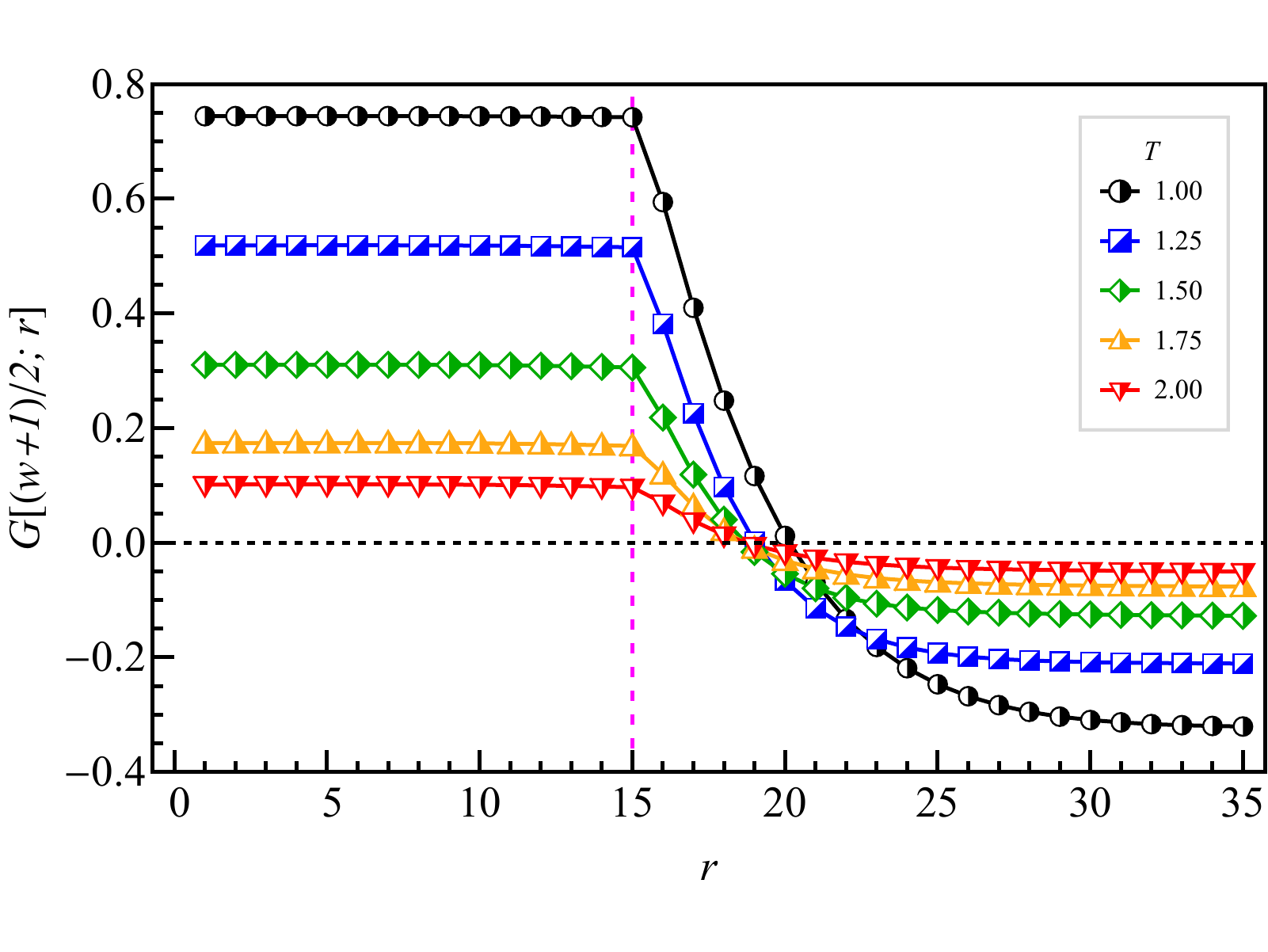}
\caption{(Color online) $G((w+1)/2,r)$ for
the two-temperature Ising ring with $L = 127$, $w = 31$, and various values
of $T$. Here the lattice site $i=(w+1)/2$ is the site in the middle of the window of width $w$.
The end of the window is indicated by the dashed vertical line.
}
\label{figure13}
\end{figure}

One of the best ways
to statistically establish the nature of the steady state is to examine the
two-spin correlations $G(i;r)$, see Eq. (\ref{Gi}). Based
on the results from the two-temperature model presented thus far, it should
be unsurprising that the corresponding correlation differs dramatically 
from that found in the standard equilibrium model, see Fig.~\ref{figure5}.
Since the translational
invariance inherent in the equilibrium model is broken by the inclusion of
the second temperature domain, it becomes necessary to specify the reference
lattice site $i$. 
In the case of Fig. \ref{figure13} the center
of the window, located at $i=(w+1)/2$, is chosen as this reference point.

The first feature of note is that $G((w+1)/2;r)$ is almost perfectly constant over the
entirety of the window. The value of this plateau agrees very well with
$\left< M_w^2 \right>/w^2$ and decreases 
for increasing $T$, thereby serving as an indicator of the frequency of
oppositely aligned spins entering the window.
Fig. \ref{figure13} also shows that in the region immediately outside of the window
the correlation decays exponentially and roughly fits the exponential behavior
of (\ref{eq:G_linear}) with different constraints.

\begin{figure}[h!]
\centering
\includegraphics[scale=.80]{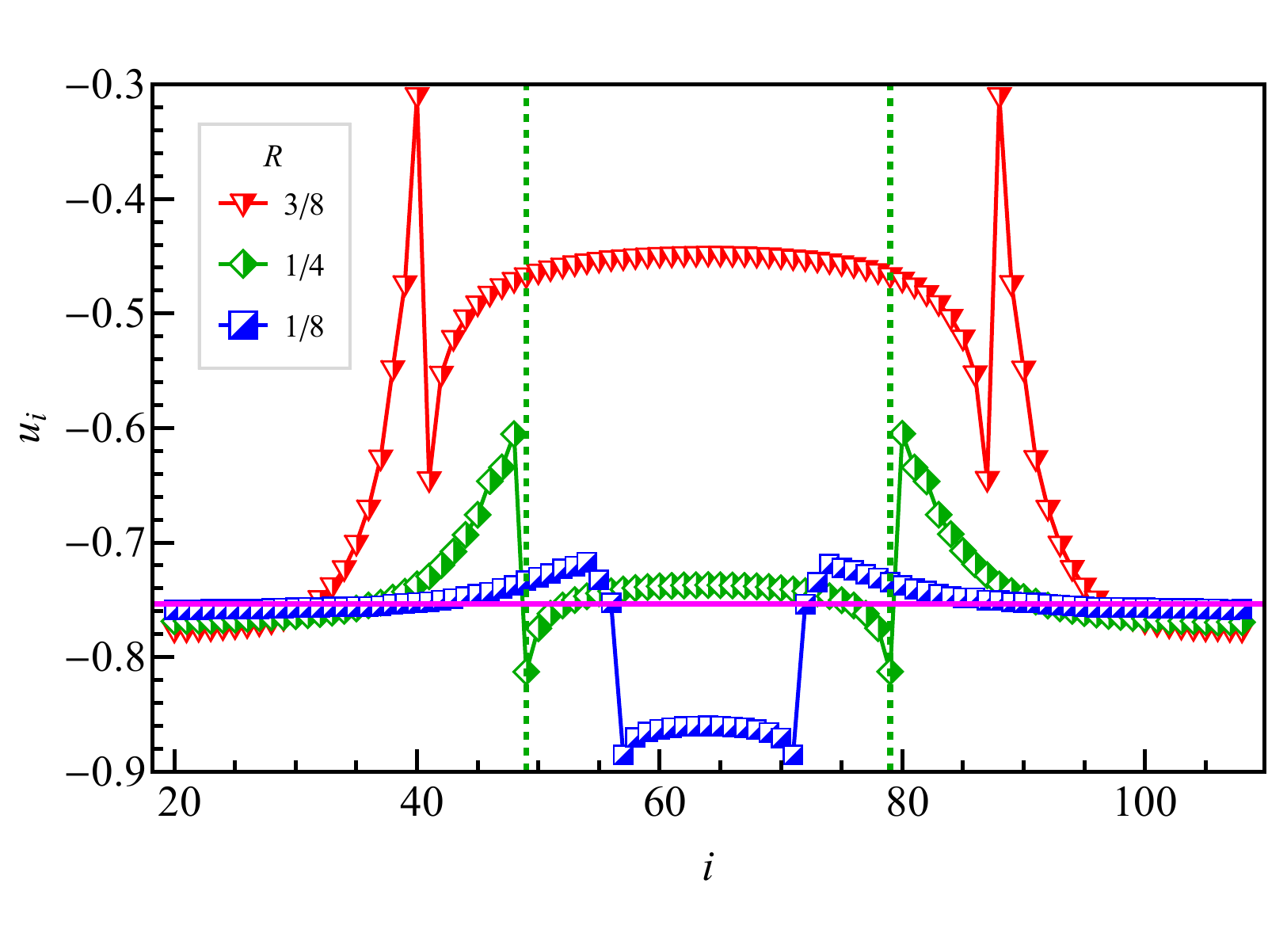}
\caption{(Color online) Average bond energy $u_i$ at site $i$ for $L = 128$, $T = 1$, and various 
ratios $R=w/L$. The solid magenta (gray) horizontal line indicates the equilibrium energy of the standard model at $T=1$.
The dashed vertical lines are placed to demonstrate the end of the window region for the $R = 1/4$ curve.
For all curves the borders of the infinite temperature window are marked by minima in $u_i$ followed by discontinuous
transitions to maxima outside the window. }
\label{figure14}
\end{figure}

Further information about the nature of the steady state
can be obtained from the average energy of the bonds
between lattice sites $u_i$, see Eq. (\ref{u}).
Displayed in Fig. \ref{figure14},
this quantity reveals the pronounced effect the boundary between
the two regions has. The equilibrium energy of the standard Ising model, indicated by the dashed horizontal line,
is approached by the 
two-temperature curve far from the window, which indicates that the two-temperature model
behaves more or less as the equilibrium system deep within the primary
lattice. As the boundary between the two temperature regions is approached,
however, the true influence of the point of broken detailed balance is
displayed. There is a sharp, discontinuous jump at the boundary between the
regions, with the average bond energy much higher immediately outside the
window than inside. This indicates that
the lattice sites immediately outside the window are notably more disordered
than those immediately inside the boundary. 
This is
true even for larger window sizes in which the energy inside the window is significantly 
higher than the equilibrium bond energy.

\begin{figure}[h!]
\centering
\includegraphics[width=0.80\columnwidth]{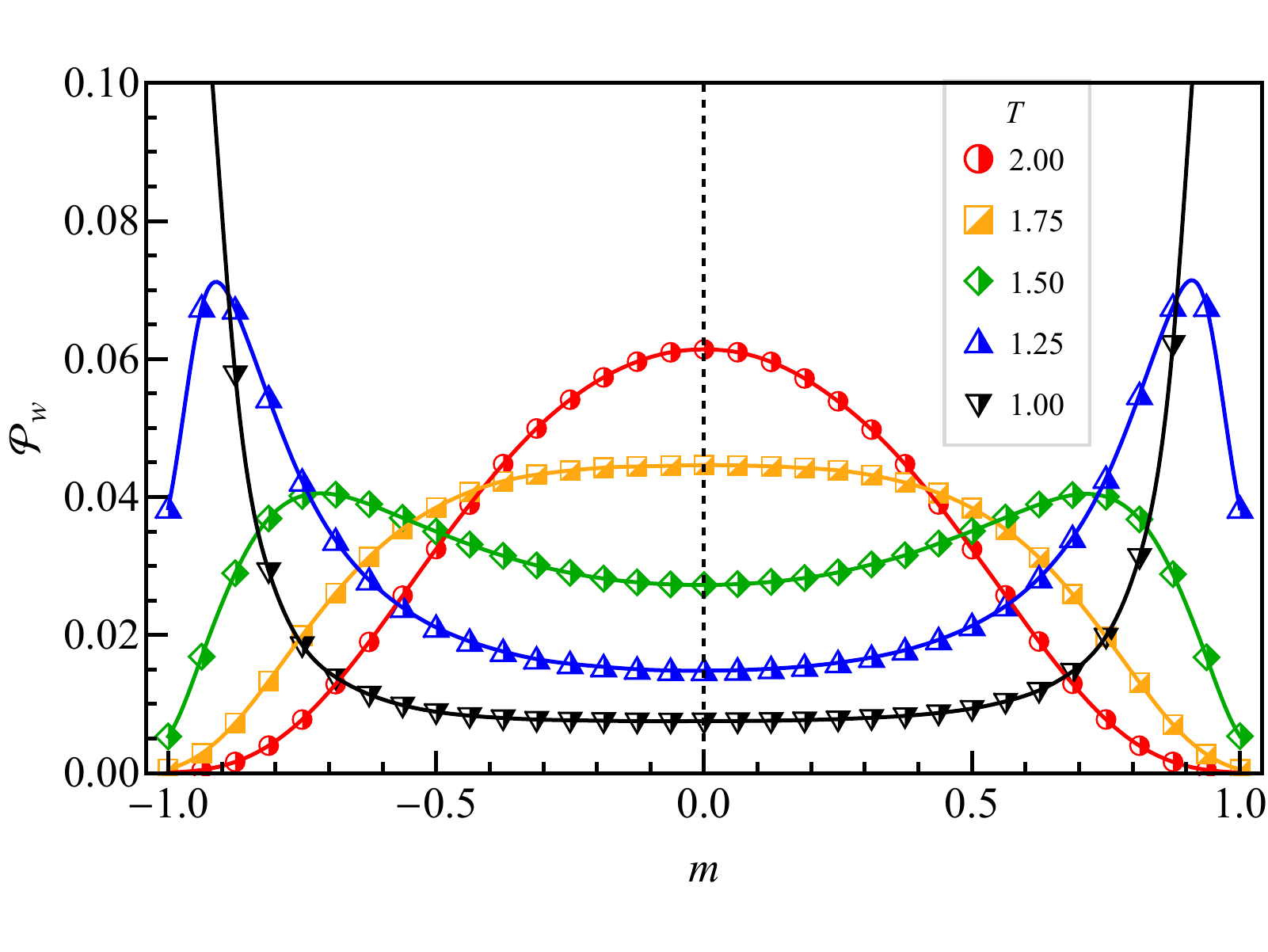}
\caption{(Color online) Distributions for the normalized window magnetization $m$
with system parameters $L = 128$ and $R=1/4$ ($w = 32$). Results for different
values of $T$ are shown.}
\label{figure15}
\end{figure}

As the final, and perhaps strongest, characterization of the nature of the
two-temperature ordered state Fig. \ref{figure15} displays the steady-state
distribution ${\mathcal P}_w(m)$ for the normalized window magnetization $m$
as a function of $T$. This
distribution provides information on how likely a certain magnetization 
is in the steady state. At relatively high temperatures, such as 
$T = 2$, the histogram takes on the expected shape of a 
distribution centered around $m = 0$, indicative of a disordered
configuration and similar to the shape observed for the standard equilibrium system,
see Fig. \ref{figure5}, as well as for the $2J$ model, see Fig. \ref{figure8}. 
As the temperature is lowered, however, there appears
to be a smooth inversion of the distribution such that half-filled states
become increasingly unlikely while wholly filled or empty states dominate.
As a result ${\mathcal P}_w(m)$ becomes bi-modal with maxima at some 
temperature-dependent normalized magnetizations $\pm m$.
The seemingly continuous change of the distribution as the temperature is
lowered is reminiscent of the behavior expected 
for a continuous phase transition across some critical temperature $T_c$. An
example of this can be found, of course, in the standard equilibrium Ising model in dimensions
$d \geq 2$, where the distribution shifts in a similar manner from a distribution
centered at zero magnetization towards a bi-modal distribution with maxima at
the spontaneous magnetization \cite{Bind81}.

%

\begin{figure}[h!]
\centering
\includegraphics[scale=.80]{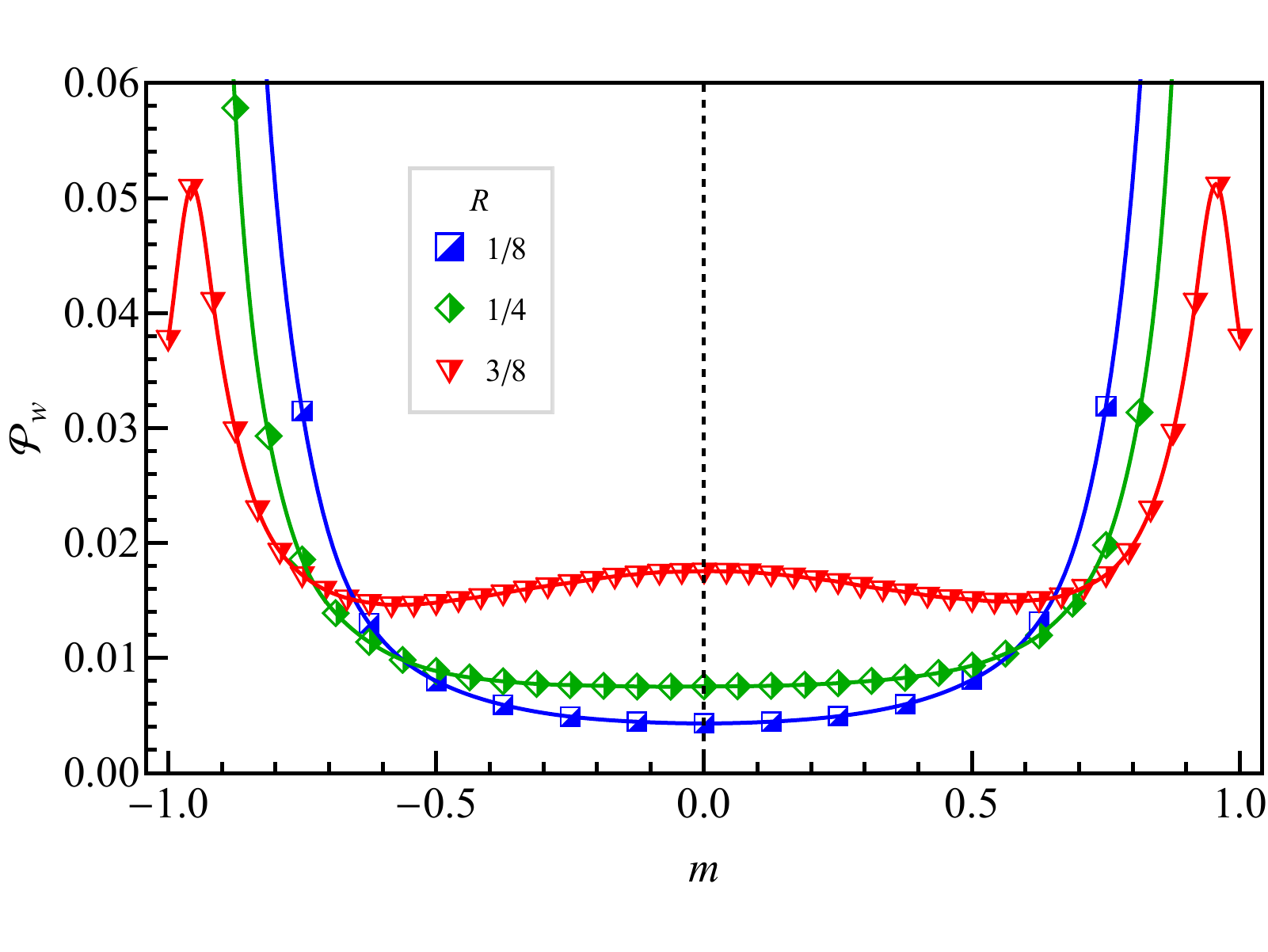}
\caption{(Color online) Distributions for the normalized window magnetization $M%
_w $ with $L = 128$ and $T = 1$ and various window sizes.}
\label{figure16}
\end{figure}

The lattice temperature $T$ is, however, not the only parameter that may
be varied to produce an apparent transition. Fig. \ref{figure16} shows that the
distribution changes considerably as the ratio $R=w/L$ is changed with constant $L$. This
happens in a similar fashion to the above changes with respect to
temperature, with the notable difference that the distribution with the maximum at $%
M = 0$ never materializes for the window sizes considered.
Instead, for larger window sizes a small central peak appears.
This could be an indication of phase coexistence.

These results indicate that the crossover between different steady states observed 
in the non-equilibrium two-temperature model has some of the finite-size hallmarks of a 
true phase transition. Whereas from the temperature-dependence one might conclude that the
transition is continuous, conflicting information is obtained when varying the window size,
with the appearance of additional peaks indicative of a system with phase coexistence. 
It is beyond the scope of the present article to further clarify the nature of this
transition, as much larger system sizes as well as other quantities need to be studied 
in order to be able to make more definitive statements. We plan to explore the nature of
this transition in more detail in the future.

\section{Conclusion}

Gaining a comprehensive understanding of non-equilibrium processes remains
an enormous task, due to the wealth and diversity of phenomena that emerge
far from equilibrium. Far from having a general theoretical framework, we
proceed by gathering valuable insights through the detailed analysis of
simple model systems.

In this paper we studied a version of the conserved one-dimensional Ising
model on a ring with minimal breaking of detailed balance. In the two
temperature model, we couple two sectors of the system to different heat
baths: one with infinite temperature for exchanges within a window of size 
$w$, and one at some temperature $T$ for the remainder of the ring. Since
spin exchanges within $w$ are indistinguishable from an Ising model with 
$J=0$ coupled to the same $T$, we also considered this two-$J$ model, in 
\textit{thermal equilibrium}. The difference between these models lies in detailed
balance being violated in the former, but only for two pairs of exchanges at the
`window edges.' This minimal modification at the interfaces entails huge
changes in the physical properties of the system, yielding a remarkable
`freezing-by-heating' ordered state. Using extensive numerical simulations
we characterized this state through a variety of quantities, most notably
the spin-spin correlations as well as the probability distribution of the
normalized window magnetization. Varying the temperature $T$ or the window
size yields a transition between a state with a highly ordered window and a
state where the window is disordered. The finite-size signatures of this
transition are ambiguous, and more work needs to be done to clarify the
nature of this `transition.' Indeed, we are aware that long range
correlations are well known to emerge in such driven diffusive systems
(breaking detailed balance in Kawasaki dynamics \cite {SZ95}), and so, 
the phenomenon
observed here may be due to finite size effects. In particular, it is
reasonable to conjecture that, if two semi-infinite systems coupled to
different baths were joined at one point, the behavior far in the bulk of
either system would be `normal,' while long range correlations induce
extended boundary layers on either side of the junction. On an extremely
large ring (with fixed $w/L$), there would be two such junctions, but widely
separated. Yet, when the ring size is small enough, the two boundary layers
will `interact' and may produce the ordered states observed here.
Substantial progress in our understanding of these systems therefore relies
on gaining a full understanding of how long range correlations can lead to
apparent long range order. In all cases, we believe that further work on
simple model systems such as this one is likely to produce novel and
surprising phenomena, as well as to provide insights into how an overarching
framework for non-equilibrium statistical mechanics can be established.

\appendix


\section{Detailed balance violation}

\begin{figure}[h!]
\centering
\includegraphics[scale=.25]{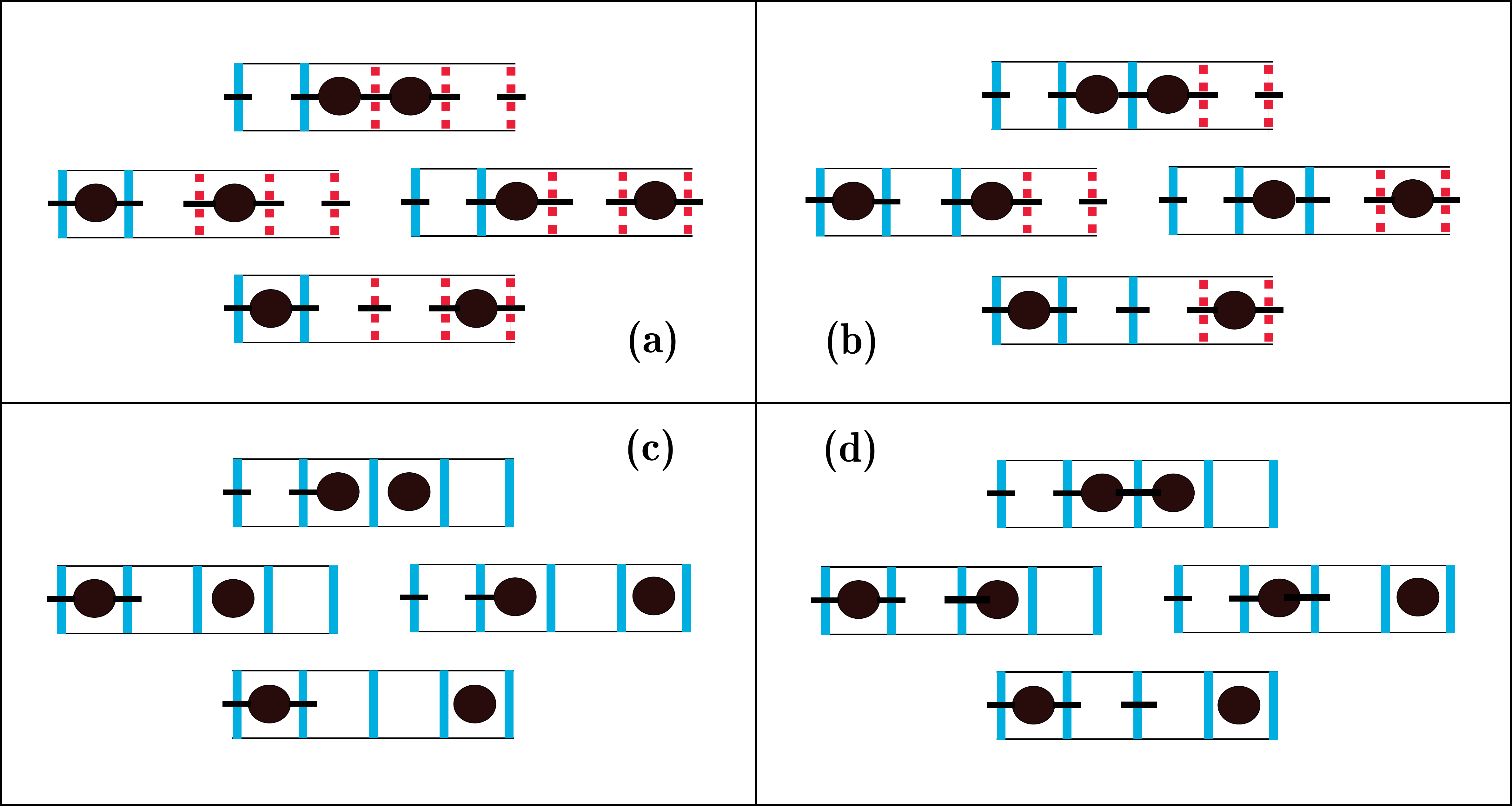}
\caption{(Color online) Two cycles of configurations illustrating irreversibility
and detailed balance violation in the $2T$ model (a,b). By comparison, these cycles 
are reversible in the $2J$ model (c,d). See text for explanation of symbols.}
\label{figure17}
\end{figure}

In this Appendix, we emphasize the crucial difference between the $2T$ and
the $2J$ model. In particular, we will illustrate detailed balance violation
in the former with \textquotedblleft irreversible\textquotedblright\ cycles
of configurations (Fig. \ref{figure17}a,b). By contrast, the same cycles in 
the $2J$ model is reversible (Fig. \ref{figure17}c,d). To make this distinction 
clear, we added a notation for the bond energies here, i.e., presence or absence 
of a horizontal dash referring to the ordinary ferromagnetic $J=1$ or the free 
$J=0$ interactions between nearest neighbors, respectively. Meanwhile, the 
vertical lines between neighbors carry the same meaning as in previous sections,
i.e., exchanges across solid (blue online) lines being updated with the finite 
$T$ and those across dashed (red online) lines being updated with $T=\infty $. 

Consider first the configurations in Fig. \ref{figure17}a, arranged around 
the quarters of a clock face. For simplicity, suppose there are no particles 
outside the four sites shown. For the clockwise cycle, using the rates for the 
$2T$ model (exchanges across different vertical lines controlled by thermal 
baths at different temperatures), the product is simply 
$1\times 1\times 1\times 1=1$. For the reverse, the rates are 
$e^{-1/T}\times 1\times 1\times 1$, which is not unity except for $T=\infty $. 
The same inequality holds for the cycles involving the configurations in 
Fig. \ref{figure17}b. The inequality of these products is a hallmark of 
detailed balance violation \cite{Kolmogorov} and signals a presence of a 
non-equilibrium stationary state \cite{ZS07}. By stark contrast, it is 
straightforward to check that the products around the cycles in 
Figs. \ref{figure17}c,d, are the same as the products for the reversed cycles. 
Of course, this equality necessarily holds for a system in thermal equilibrium,
governed by the Hamiltonian $ -J\Sigma _{i}\sigma _{i}\sigma _{i+1} $
(with appropriate $J=1,0$).


\section{Equilibrium Ising model}

The one-dimensional Ising model (on a ring or with open boundaries), solved
in 1925 \cite{Ising}, appears in most textbooks on statistical mechanics. We
summarize some of its properties here for the convenience of the reader. In
particular, most of the texts deal with the simplest case (thermodynamic limit
and no constraint on the total magnetization). Since we are not aware of any
literature that displays the explicit form of say, the correlation function for a
finite system with fixed $M$, we present some of its behavior here.

The partition function of the system of interest here is 
\begin{equation}
Z\left( T;M,L\right) =\sum_{\left\{ \sigma _{i}\right\} }\delta \left(
M-\sum\limits_{i=1}^{L}\sigma _{i}\right) \exp \left\{ K\sum\limits_{i=1}^{L}\sigma
_{i}\sigma _{i+1}\right\}
\end{equation}%
where $K\equiv J/k_{B}T$. The constraint prevents a simple evaluation of the
configuration sum, but its generating function%
\begin{equation}
\Omega \left( M\right) \equiv \sum_{M}z^{M}Z\left( T;M,L\right)
\end{equation}%
can be found easily with the transfer matrix:%
\begin{equation}
{\mathbb T}\left( z,K\right) \equiv \left( 
\begin{array}{cc}
ze^{K} & ze^{-K} \\ 
e^{-K}/z & e^{K}/z%
\end{array}%
\right)
\end{equation}%
so that%
\begin{equation}
\Omega =\mbox{Tr}\,{\mathbb T}^{L}
\end{equation}%
Then, $Z$ can be found through the integral%
\begin{equation}
\frac{1}{2\pi i}\oint \frac{dz}{z^{M+1}}\left( \lambda _{+}^{L}+\lambda
_{-}^{L}\right)  \label{I}
\end{equation}%
where%
\begin{equation}
\lambda _{\pm }=e^{K}\left[ \cosh H\pm \sqrt{\sinh ^{2}H+e^{-4K}}\right]
\end{equation}%
are the eigenvalues of ${\mathbb T}$, with $H\equiv \ln z$. Of course, saddle
point methods can be exploited to give us $Z$ as an asymptotic expansion in
large $L$.

For the two point correlation function, $G\left( r\right) \equiv
\left\langle \sigma _{i}\sigma _{i+r}\right\rangle $, we have the exact
expression%
\begin{equation}
G\left( r\right) =Z^{-1}\sum_{\left\{ \sigma _{i}\right\} }\sigma _{L}\sigma_r
\delta \left( M-\sum\limits_{i=1}^{L}\sigma _{i}\right) \exp \left\{ K\sum\limits
_{i=1}^{L}\sigma _{i}\sigma _{i+1}\right\}
\end{equation}%
which can be found through 
\begin{equation}
\mbox{Tr}\left\{ \left( 
\begin{array}{cc}
1 & 0 \\ 
0 & -1%
\end{array}%
\right) {\mathbb T}^{r}\left( 
\begin{array}{cc}
1 & 0 \\ 
0 & -1%
\end{array}%
\right) {\mathbb T}^{L-r}\right\}
\end{equation}%
and another, somewhat more involved, integral. Instead of pursuing these
exact expressions, it is reasonable to appeal to an approximation, which
should be quite good here. We start with the exact expression
\begin{equation}
G_{0}(r)=\left( \omega ^{r}+\omega ^{L-r}\right) /\left( 1+\omega ^{L}\right)
\end{equation}%
where $\omega \equiv \tanh K$, for a finite ring {\it without} constraints
on $M$. Assuming a linear relationship between $G_{0}$ and our $G$, we fix
the constants by imposing $G\left( 0\right) =1$ and the constraint%
\begin{equation}
\sum_{r=0}^{L-1}G\left( r\right) =M^{2}/L
\end{equation}%
(a consequence of $M=\Sigma _{i=1}^{L}\sigma_i $). The result is 
\begin{equation}
G\left( r\right) \cong \frac{G_{0}\left( r\right) -B}{1-B}
\end{equation}%
with 
\begin{equation}
B=\frac{1}{L^{2}-M^{2}}\left[ L\frac{\left( 1-\omega ^{L}\right) \left(
1+\omega \right) }{\left( 1+\omega ^{L}\right) \left( 1-\omega \right) }%
-M^{2}\right]
\end{equation}%
Thus,%
\begin{equation}
G\left( r\right) \cong \frac{\left( \omega ^{r}+\omega ^{L-r}\right) -\left(
1+\omega ^{L}\right) B}{\left( 1+\omega ^{L}\right) \left( 1-B\right) }
\end{equation}%
which is the form (\ref{eq:G_linear}).
If we impose $M=0$ and consider $r$'s such that $\omega ^{L-r}\ll \omega
^{r} $, then 
\begin{equation}
B\rightarrow e^{2K}/L
\end{equation}%
and 
\begin{equation}
G\left( r\right) \cong \frac{\omega ^{r}-e^{2K}/L}{1-e^{2K}/L}~.
\end{equation}%

Turning to the distribution ${\cal P}_{w}\left( m\right) $, it is given by%
\begin{equation}
{\cal P}_{w}\left( m,M\right) =Z^{-1}\sum_{\left\{ \sigma _{i}\right\}
}\delta \left( m-\sum\limits_{i=1}^{w}\sigma _{i}/w\right) \delta \left(
M-\sum\limits_{i=1}^{L}\sigma _{i}\right) \exp \left\{ K\sum\limits_{i=1}^{L}\sigma
_{i}\sigma _{i+1}\right\} 
\end{equation}%
Its generating function 
\begin{equation}
{\it \Omega}_{w}\left( \zeta ;z\right) \equiv \sum_{m,M}\zeta ^{mw}z^{M}%
{\cal P}_{w}\left( m,M\right) 
\end{equation}%
can be computed as above:%
\begin{equation}
\mbox{Tr}\left[ {\mathbb T}\left( \zeta z,K\right) ^{w}{\mathbb T}\left( z,K\right) ^{L-w}%
\right] 
\end{equation}%
while inverting it to ${\cal P}_{w}$ will require two contour integrals.
Though feasible, we will not pursue these results here. Simulations provide
a more direct route to ${\cal P}_{w}$, and we are not interested in its
analytic properties.

\bigskip 

\bigskip

\begin{acknowledgments}
We thank B. Schmittmann for contributions in the initial stages of this
research. One of us (RKPZ) is grateful to B. Derrida, H.W. Diehl, E. F. Redish, and R. Swendsen for
illuminating discussions. M.P. and R.K.P.Z. thank the INFN and the Galileo
Galilei Institute for Theoretical Physics for hospitality and
for partial support. This work is supported in part by the US National
Science Foundation through grants DMR-1205309 and DMR-1244666.
\end{acknowledgments}


\begin{thebibliography}{99}
\bibitem{footnote} In the following we express energies in units of $J$,
whereas temperatures are expressed in units of $J/k_B$.
\end{thebibliography}

\end{document}